\documentclass[]{aa}  

\usepackage{graphicx}
\usepackage{txfonts}
\usepackage{hyperref}
\newcommand{\review}{}
\newcommand{\reviewtwo}{}

\begin{document}

   \title{Improving the open cluster census.}

   %DIFDELCMD < \subtitle{I. Comparison of clustering algorithms applied to Gaia DR2 data.}
%DIFDELCMD < %%%
\subtitle{I. Comparison of clustering algorithms applied to \emph{Gaia} DR2 data.}

   \author{Emily L. Hunt\inst{1}\fnmsep\thanks{Fellow of the International Max Planck Research School for Astronomy and Cosmic Physics at the University of Heidelberg (IMPRS-HD).}
           \and
           Sabine Reffert\inst{1}
          }

   \institute{Landessternwarte, Zentrum für Astronomie der Universität Heidelberg, Königstuhl 12, 69117 Heidelberg, Germany\\
              \email{ehunt@lsw.uni-heidelberg.de}
             }

   \date{Received 4 September 2020 / accepted ???}

% \abstract{}{}{}{}{} 
% 5 {} token are mandatory

  \abstract
   {The census of open clusters in the Milky Way is in a never-before seen state of flux. Recent works have reported hundreds of new open clusters thanks to the incredible astrometric quality of the \emph{Gaia} satellite, but other works have also reported that many open clusters discovered in the pre \emph{Gaia} era may be associations.}
   {We aim to conduct a comparison of clustering algorithms used to detect open clusters, attempting to statistically quantify their strengths and weaknesses by deriving the sensitivity, specificity, and precision of each as well as their true positive rate against a larger sample.}
   {We selected DBSCAN, HDBSCAN, and Gaussian mixture models for further study, owing to their speed and appropriateness for use with \emph{Gaia} data. \review{We }developed a preprocessing pipeline for \emph{Gaia} data and developed the algorithms further for the specific application to open clusters\review{. We derived detection }rates for all \review{1385 }open clusters in \review{the }fields in our study \review{as well as more detailed performance statistics for 100 of these open clusters}.}
   {DBSCAN was sensitive to 50\% to 62\% of the true positive open clusters in our sample, with generally very good specificity and precision. HDBSCAN traded precision for a higher sensitivity of \review{up to }82\%, especially across different distances and scales of open clusters. Gaussian mixture models were slow and only sensitive to 33\% of open clusters in our sample, which tended to be larger objects. \review{Additionally, we report on 41 new open cluster candidates detected by HDBSCAN, three of which are closer than 500~pc.}}
   {When used with additional post-processing to mitigate its false positives, we have found that HDBSCAN is the most sensitive and effective algorithm for recovering open clusters in \emph{Gaia} data. \review{Our results suggest that }many more new and already reported open clusters have yet to be detected in \emph{Gaia} data.}


   \maketitle
%
%-------------------------------------------------------------------

\section{Introduction}

Open clusters (OCs) are commonly known as the laboratories of stellar evolution, which form when large gas clouds collapse into dense, gravitationally bound regions of stars. The stars in OCs have roughly the same age and chemical composition, meaning that every OC is a unique `experiment' showing the results of stellar evolution with stars across a range of masses given a certain set of initial conditions. In particular, OCs in our own galaxy are the most enlightening to study, since their proximity means that individual stars can be resolved and parameters can be determined to higher levels of precision.

The number of known open clusters has not changed significantly until recently. The New General Catalogue (NGC) listed $\approx$700 objects that we now know to be OCs \citep{dreyer_new_1888}, the most comprehensive catalogue of its time -- yet over a century later, the catalogue of \cite{mermilliod_database_1995} had only increased to a size of ~1200 OCs, not even doubling the OC census despite the large strides in astronomical instrumentation and data analysis taken in the 20th century. In part, this is because numerous clusters in the literature were ruled out as associations by modern data, reducing the size of the census -- yet it still persists that a century of work did not significantly increase the size of the OC census.

The largest increases to the size of the census came with the advent of new techniques. The space-based astrometric survey of the \emph{Hipparcos} satellite \citep{hog_tycho-2_2000} revealed a number of new, often relatively sparse OCs in studies such as \cite{platais_search_1998} and \cite{chereul_distribution_1999}, while wide-field infrared surveys looked through interstellar extinction to find new OCs in studies such as \cite{dutra_new_2001} and \cite{froebrich_systematic_2007}. The catalogue of \cite{kharchenko_global_2013} (hereafter MWSC) lists 2267 probable OCs and a further 132 that showed nebulosity, a major increase from the figure of \cite{mermilliod_database_1995} just two decades prior.

The next major increase to the size of the OC catalogue is currently in progress thanks to the \emph{Gaia} satellite \citep{brown_gaia_2018}. \emph{Gaia} maps the stars of the Milky Way in five dimensions (positions, proper motions, and parallax), while also providing visual photometry and colours in its own $G$, $G_{BP}$, and $G_{RP}$ photometric bands, and spectroscopic radial velocities for a small sample of bright stars. Compared with \emph{Hipparcos}, \emph{Gaia} has roughly an order of magnitude more precision in astrometric parameters for $10^4$ times as many stars, resulting in a groundbreaking dataset that has full astrometric solutions and photometry for 1.3 billion stars as of \emph{Gaia} DR2, 7 million of which also have radial velocities.

It is perhaps unsurprising that such a large improvement in our ability to map the galaxy is also greatly improving the OC census. In terms of quantity, works such as \cite{castro-ginard_new_2018,castro-ginard_hunting_2019,castro-ginard_hunting_2020}, \cite{liu_catalog_2019}, \cite{sim_207_2019}, and \cite{cantat-gaudin_gaia_2019} have recently reported hundreds of candidate OCs using \emph{\review{Gaia}} data. Typically, this is done using automated blind searches of the \emph{\review{Gaia}} dataset with clustering algorithms -- a type of unsupervised machine learning that can find the most natural groupings or clusters within a dataset, requiring only basic parameters and minimal prior knowledge about the structure of the data.

The precision of \emph{Gaia} is also improving the quality of the OC census. While traditionally, distances to OCs would be derived using photometry alone and fitting a model-dependent stellar isochrone to the OC's colour-magnitude diagram, \emph{Gaia} parallaxes provide an unbiased and model-independent distance estimator -- allowing parameters for OCs to be derived to greater levels of precision \review{\citep[e.g.][]{cantat-gaudin_painting_2020}
}. \cite{cantat-gaudin_gaia_2018} derive membership lists and parameters for 1229 OCs using \emph{Gaia} DR2 data alone, which has been expanded with some re-analysis and by including recently detected clusters in \cite{cantat-gaudin_clusters_2020}. It is expected that some OCs listed in MWSC will not be detectable in \emph{Gaia} data. \emph{Gaia}'s visual band observations are unable to see into areas of high dust extinction unlike the infrared photometry used by MWSC -- obscuring small, distant OCs from view in regions with high extinction, such as towards the galactic centre at distances greater than $\sim3$ kpc. In addition, parallaxes and proper motions have fractional uncertainties that increase with distance, which has a significant negative effect on the signal to noise ratio of OCs in \emph{Gaia} data at distances larger than $\sim1-3$ kpc.

However, despite astrometric uncertainties or dust obscuring \emph{Gaia}'s view of some clusters, it is also possible to rule out a number of OCs that should still be detectable in \emph{Gaia} data based on their existing parameters. \cite{cantat-gaudin_clusters_2020} have ruled out 38 OCs in the literature as asterisms, all of which should be bright enough to detect in the \emph{Gaia} dataset based on their reported parameters but do not appear to exist. Future studies will be able to rule out yet more putative OCs based on \emph{Gaia} data alone, particularly as \emph{Gaia} data improves in the coming years with future releases.

In its current state, the OC census is difficult for astronomers to use. Despite MWSC deriving that the OC census was complete to within 1.8 kpc, the recent myriad of studies using \emph{Gaia} data have shown that many more OCs are yet to be discovered within the immediate solar neighbourhood. Until the OC census is shown to be complete to within a certain radius, it is impossible to calculate accurate population statistics about OCs in the Milky Way. In addition, the many asterisms that are not yet concretely ruled out in the literature make the OC census more difficult to use, as not all reported OCs in the literature are really there and many do not make good targets for precious telescope time.

Within the next decade, future data releases of the \emph{Gaia} satellite and large-scale spectroscopic surveys such as 4MOST \citep{de_jong_4most_2012} will provide astronomers with a wealth of data on on our galaxy. OCs are an important piece of the jigsaw puzzle of the Milky Way's current and past star formation. An OC census with greatly improved quality and quantity will allow astronomers to use the census reliably for a \review{range }of scientific purposes, including mapping the age distribution of OCs across the galaxy \citep{cantat-gaudin_painting_2020, yen_reanalysis_2018}, studies of the chemical composition of OCs \citep{baratella_gaia-eso_2020, donor_open_2020}, to even studying the conditions of planet formation in OCs and the implications that may have for the distribution of the wider exoplanet census \citep{fujii_survival_2019}.

To date, a number of different methods have been used to search for new or existing OCs in \emph{Gaia} data. While UPMASK \citep[Unsupervised Photometric Membership Assignment in Stellar Clusters,][]{krone-martins_upmask:_2014} as used by \cite{cantat-gaudin_gaia_2018} and \cite{cantat-gaudin_clusters_2020} is a highly successful tool for producing membership lists of existing OCs, it is too slow to conduct a large-scale blind search across the billion star dataset of \emph{Gaia}. In turn, while approaches such as the one applied in \cite{castro-ginard_hunting_2020} has detected hundreds of new OCs in the \emph{Gaia} dataset, their method is unable to detect a large fraction of literature OCs, suggesting that their approach may also be unable to detect a large fraction of as yet undiscovered OCs with similar properties. Different approaches have advantages and disadvantages that have never before been compared side-by-side on \emph{Gaia} data, and no single approach has yet been developed that can simultaneously detect new OCs in a large-scale blind search while also detecting a majority of already-reported objects.

In this series of papers, we will work to improve the OC census: primarily by attempting to detect new OCs, but also by re-detecting a large fraction of literature OCs with a different methodology and complementing cataloguing efforts such as \cite{cantat-gaudin_clusters_2020}. In this study, we create an unbiased preprocessing pipeline to prepare \emph{Gaia} data for analysis by clustering algorithms, and test the ability of three clustering algorithms to detect OCs in \emph{Gaia} data. In Sect.~\ref{sec:data}, we describe the \emph{Gaia} data used and the applied pre-processing steps. Section~\ref{sec:algorithms} outlines the requirements for any clustering algorithm to be applied to \emph{Gaia} data and describes three chosen algorithms that meet these criteria. Our analysis process for the algorithms applied to our data and the results of this are presented in Sect.~\ref{sec:results}. In Sect.~\ref{sec:discussion}, we discuss the strengths and weaknesses of each approach and the implications for future studies. \review{We report on 41 new OC candidates discovered during the preparation of this paper in Sect.~\ref{sec:new_ocs}. Finally, Sect.~}\ref{sec:conclusion} summarises our results.

%--------------------------------------------------------------------

\section{Data}\label{sec:data}
\subsection{The \emph{Gaia} DR2 dataset and the HEALPix system}

\begin{figure*}
   \centering
   \includegraphics{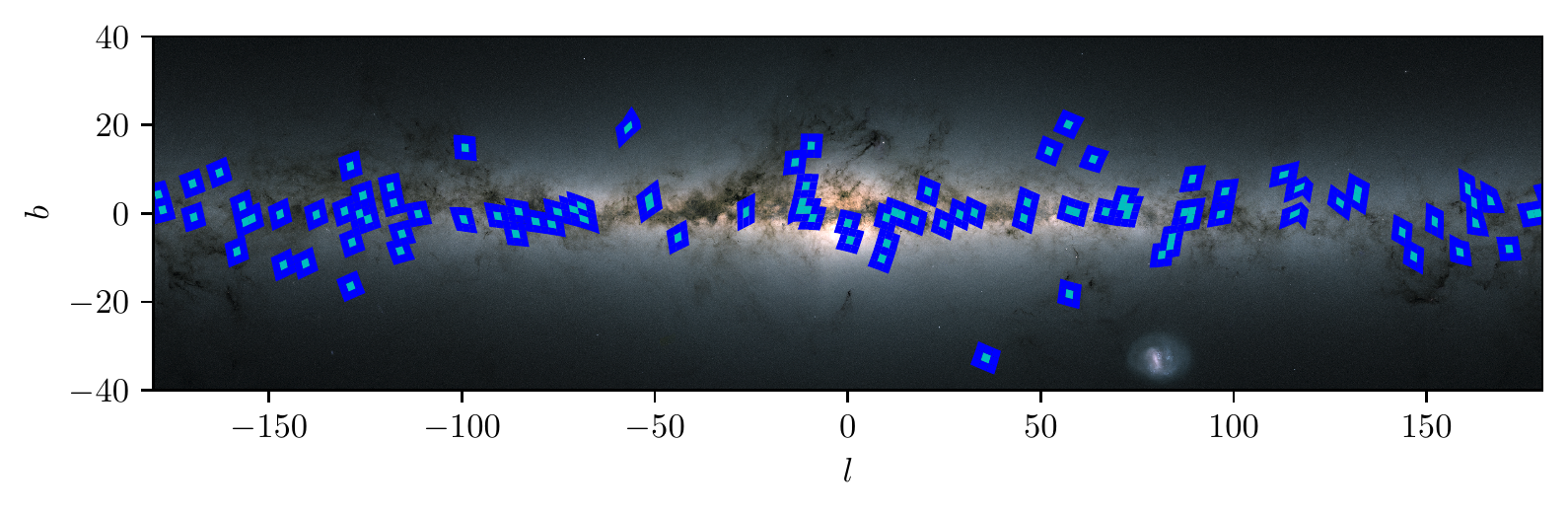}
   \caption{Target fields for this study plotted above a  \emph{\review{Gaia}} map of stellar density in an equirectangular projection. Cyan regions show the 100 main HEALPix level five pixels. Each main pixel was merged with its eight nearest neighbours, which are shown in blue. Some nearest neighbour pixels overlap between different fields.}\label{fig:targetfields}%
\end{figure*}

The results of any unsupervised search for OCs are always highly dependent on the input data and how it is preprocessed: assumptions must be made for reasons of computational efficiency (for instance, splitting the dataset into separate chunks to improve runtime), and dimensions of the data with different units and coordinate systems must be intelligently preprocessed to allow an unsupervised algorithm to take full advantage of \emph{Gaia} data. We briefly introduce the \emph{Gaia} satellite and explain the preprocessing pipeline we developed to prepare its data for use with unsupervised clustering algorithms.

The \emph{Gaia} satellite is producing a previously unprecedented quantity and quality of astrometric and photometric data for stars in the Milky Way. 1.7 billion sources brighter than $G=21$ are included in \emph{Gaia} DR2, where 1.3 billion have full five-parameter astrometric solutions. Uncertainties on derived parameters for each source depend strongly on the brightness of the source. While as many detected sources as possible are included in \emph{Gaia} DR2 for completeness, the majority of faint sources are not useful for studies of galactic structure as the uncertainties on their parameters are too large. 

For example, a star with brightness $G=17$ would have corresponding uncertainties of 0.1 mas in parallax and 0.2 mas yr$^{-1}$ in proper motion \citep{brown_gaia_2018}. If this star is 1 kpc away and hence has a true parallax of 1 mas, a measured parallax for this star would be informative to within roughly 10\% of the true distance. The uncertainty on proper motion for this star would easily allow it to be distinguished as a member of an open cluster, as open clusters at this distance typically have an inherent proper motion dispersion of $\sim$1 mas yr$^{-1}$ which is larger than the star's proper motion uncertainty. However, a faint star with $G=20$ at a true distance of 1 kpc will have corresponding uncertainties of 0.7 mas in parallax and 1.2 mas yr$^{-1}$ in proper motion. Any parallax measurement for this star will be much less informative about the star's true distance, with a near 100\% fractional uncertainty. Its proper motion uncertainty is larger than the typical dispersion of proper motion in OCs at 1 kpc, meaning that this faint star could never be reliably assigned as a member of an OC with \emph{Gaia} DR2.

As such, most studies adopt a cut on the dataset to ignore uninformative stars and improve the signal to noise ratio of open clusters in the \emph{Gaia} data. For the purposes of this study, we cut all stars fainter than $G=18$, corresponding to typical maximum uncertainties of 0.15 mas in parallax and 0.3 mas $yr^{-1}$ in proper motion, which is the same magnitude cut as used by \cite{cantat-gaudin_gaia_2018} and \cite{liu_catalog_2019}, although \cite{castro-ginard_hunting_2020} adopt a stronger cut at $G=17$.

Some studies \citep[such as][]{castro-ginard_hunting_2020, liu_catalog_2019} also remove outlier stars based on the magnitude of their proper motions or parallaxes. We choose not to remove stars with negative parallaxes, as this would make our study less sensitive to the most distant clusters for which member stars may have zero or negative parallax values. We also do not remove stars with high proper motions -- while very few open clusters have proper motions $\mu_{\alpha*}$ or $\mu_\delta$ of greater than 30 mas yr$^{-1}$, we still wish to have as few biases as possible in this study.

To select regions of the sky for study, we use the HEALPix\footnote{\href{http://healpix.sourceforge.net}{http://healpix.sourceforge.net}} (Hierarchical Equal Area isoLatitude Pixelization) scheme \citep{gorski_healpix:_2005} to select equal-area approximately quadrilateral regions of the \emph{\review{Gaia}} dataset. HEALPix has advantages over rectangular tesselation schemes \citep[such as those used in][]{castro-ginard_hunting_2020} since spherical distortions from projecting quadrilaterals onto the sky are spread out, allowing algorithms to be ran on equal-area pixels. In addition, \emph{Gaia} sources are numbered based on a HEALPix system, making the HEALPix system convenient for a study of \emph{Gaia} data to implement.

We aim to tile the sky into manageable chunks: large enough to contain OCs beyond a certain distance, but not so large that the amount of data in each chunk becomes prohibitively computationally expensive to run on. These chunks should also be easy to overlap, so that our future blind searches would not have edge effects. To do this, we select a region of study and its HEALPix level five pixel. The eight nearest pixels to each central pixel are added to each region of data to analyse, creating chunks each of area $\sim 31$ deg$^2$ and side length approximately 5$^\circ$. 0.2\% of HEALPix pixels only have seven neighbours and will hence have slightly smaller areas. It may be difficult to detect OCs closer than $\sim$350 pc with this tiling scheme since a typical OC at this distance may have a larger tidal radius that the data field. Any future blind search would be supplemented with a clustering analysis in Cartesian co-ordinates of all stars within 500 pc, solving this issue and also allowing nearby OCs to be properly detected without issues stemming from spherical distortions at angular separations greater than roughly $\sim10^{\circ}$.

HEALPix is straightforward to use with \emph{Gaia} data, since all stars are numbered based on the HEALPix pixels they are present in. For a given \texttt{source\_id}, its HEALPix pixel at level $n$ is given by:

\begin{equation}\label{eqn:healpix}
    \textrm{HEALPix pixel} = \textrm{FLOOR} \left( \frac{\texttt{source\_id}}{2^{35} \cdot 4^{12-n}}  \right).
\end{equation}

\noindent
For efficiency when querying the \emph{Gaia} database with ADQL, this formula is inverted and used to select all stars with a \texttt{source\_id} in the correct range of values. When downloading the stars in a given pixel, we require that all sources have a full five-parameter astrometric solution, valid $G_{BP}$ and $G_{RP}$ photometry, and a $G$-band magnitude less than 18. An exact copy of the ADQL query used for this study is included in Appendix~\ref{app:adql}.

%--------------------------------------------------------------------
\subsection{Selection of target fields}

To study the effectiveness of clustering algorithms across a representative sample of stellar densities, we randomly selected 100 objects from MWSC that were each in unique HEALPix level five pixels. This list of 100 objects formed a list of 100 `main objects' to study. The eight nearest pixels to each central pixel were added to each of the 100 selected pixels to analyse. This resulted in 100 separate chunks, with 733 unique HEALPix level five pixels out of 900 in total since the neighbour pixels of different chunks were allowed to overlap. The fields are shown in Fig.~\ref{fig:targetfields} and listed in Appendix~\ref{app:fields}. \review{The fields contained between 100$\,$000 to 4.2 million stars, with a mean of 734$\,$000 stars. In total, all fields contain 56.8 million unique stars, representing $\approx$20\% of the 260 million stars in }\emph{\review{Gaia}} \review{DR2 brighter than $G=18$.
}

All but one of these fields are in the galactic disk with $|\,b\,|<25^{\circ}$, and many of them are situated in areas of dense star formation where many OCs are present. We accidentally selected two globular clusters in our main list that did not contain any OCs centrally located in their field, which we replaced in our main list of 100 OCs with OCs from fields 14 and 57. To expand the target list from the initial 100 OCs to include other OCs contained within these fields, we searched the catalogues of MWSC, \cite{cantat-gaudin_clusters_2020}, \cite{castro-ginard_hunting_2020} and \cite{liu_catalog_2019}, which contain \review{a total sum of 4002 reported OCs}. We required that reported OCs in the literature be entirely contained by a field given their reported radius and distance \review{to mitigate edge effects which could cause non-detections}. For \cite{cantat-gaudin_clusters_2020} OCs, $2 \cdot r_{50}$ (the radius containing half of the members of the OC) was used as a proxy for tidal radius. For the catalogue of \cite{castro-ginard_hunting_2020}, which \review{lists Gaussian angular dispersions $\theta$ containing $\sim68$\% of members, $2 \cdot \theta$ was used as a proxy for tidal radius}. In total, the literature reports \review{1385 }unique OCs contained within the 100 fields, all of which should be entirely visible and not partially clipped by the fields' edges. This represents roughly a third of the total number of clusters that the above four works report.

\review{The objects in MWSC that remain undetected in }\emph{\review{Gaia}} \review{data present a particular challenge for the algorithms in this study. }Since \cite{cantat-gaudin_clusters_2020} have found that a number of clusters listed in MWSC are not real, we do not expect any algorithm to detect OC candidates corresponding to all listed targets\review{, meaning that most MWSC targets are false positives that should be discarded by the algorithms. However, a small number of MWSC objects may be real but are simply as yet undetected in }\emph{\review{Gaia}} \review{data due to limitations of the methodologies used. Hence, the inclusion of MWSC objects allows us to test the ability of the algorithms to rule out putative objects, corresponding to their true and false negative rates, while also testing the algorithms against the sensitivity of existing approaches and seeing if any additional MWSC targets can be recovered in \emph{Gaia} data with new methodologies. }We chose to use real \emph{Gaia} data for our study instead of simulated data so that we can develop our full pipeline from start to finish to work with real data, which includes a number of challenging aspects \cite[such as systematic errors on astrometric parameters,][]{lindegren_gaia_2018} that would not be adequately tested by using simulated data only.

%--------------------------------------------------------------------
\subsection{Preprocessing steps}

In the limit of small errors, clustering analysis could be performed with three-dimensional spatial data in a Cartesian frame. However, parallaxes are inherently difficult to measure and have large fractional uncertainties in \emph{Gaia}, and transforming the spherical co-ordinate system of \emph{Gaia} data to Cartesian co-ordinates is non-trivial and would introduce large errors to other axes of the data. As such, it is easier to remain in a spherical co-ordinate system to avoid contaminating positional data with the large errors of parallax measurements. Searches for OCs are helped immensely by proper motions, as OCs are gravitationally bound groups of stars that appear tightly clumped in proper motion space. These could be changed to Cartesian velocities with parallaxes, but this is avoided for the same reasons as with positions. 

Attempts were made to use distances instead of parallaxes with the distance catalogue of \cite{bailer-jones_estimating_2018}. However, while their method is appropriate for macroscopic studies of galactic structure, it places stars with uncertain parallaxes at a prior-defined distance, which moves low magnitude member stars further from their parent OCs in the data and was found to reduce the signal to noise ratio of OCs in the data. \review{Alternative distance estimators that are better at preserving small scale galactic structure could be investigated in the future, }such as StarHorse \citep{anders_photo-astrometric_2019} \review{which uses }magnitude information and stellar models to increase the accuracy of \emph{Gaia}-derived distances.

Some pre-processing can be done to reduce the effect of remaining in a spherical co-ordinate system and using parallaxes instead of distances, but without contaminating other dimensions of the data with the large uncertainties of parallax measurements. To remove spherical distortions that occur at high latitudes in position and proper motions, every field is rotated to an arbitrary co-ordinate frame $(\lambda, \, \phi)$ centred at (0,0) and rotated to have edges parallel with the co-ordinate axes for neater plotting of individual fields, with proper motions $\mu_{\alpha*}$, $\mu_{\delta}$ also transformed to the new frame as $\mu_{\lambda*}$, $\mu_{\phi}$. 

Machine learning algorithms benefit from having scaled inputs, so the five dimensions of data for each field $(\lambda, \, \phi, \, \mu_{\lambda*}, \, \mu_{\phi}, \, \varpi)$ are re-scaled to have a median of zero and a unit inter-quartile range using a \texttt{RobustScaler} object from \texttt{scikit-learn} \citep{pedregosa_scikit-learn_2011}. This process is resilient to outliers, unlike scaling to have zero mean and unit variance as is sometimes used in the literature. This re-scaling process also ensures that each co-ordinate axis has an equal weight when passed to clustering algorithms. We choose not to experiment with re-weighting dimensions of the dataset as was performed tentatively by \cite{liu_catalog_2019}, although this could be explored in future works.

%--------------------------------------------------------------------

\section{Selection and implementation of clustering algorithms}\label{sec:algorithms}
\subsection{Criteria}\label{sec:algorithms_criteria}

% Algorithms table
\begin{table}
\caption{Algorithms considered for inclusion by this study.}
\label{tab:algorithms}
\centering
\begin{tabular}{l c c c}
\hline\hline
          & Runtime & Deals with & Open-  \\
Algorithm & scaling\tablefootmark{a} & noise      & source \\
\hline                        
KMeans                & $n$        & No  & \texttt{sklearn}\tablefootmark{b} \\
Affinity propagation  & $n^2$      & No  & \texttt{sklearn}\tablefootmark{b} \\
Mean-shift            & $n^2$      & No  & \texttt{sklearn}\tablefootmark{b} \\
Spectral              & $n^3$      & No  & \texttt{sklearn}\tablefootmark{b} \\
Ward                  & $n^3$      & No  & \texttt{sklearn}\tablefootmark{b} \\
Agglomerative         & $n^3$      & No  & \texttt{sklearn}\tablefootmark{b} \\
DBSCAN                & $n \log n$ & Yes & \texttt{sklearn}\tablefootmark{b} \\
OPTICS                & $n^2$      & Yes & \texttt{sklearn}\tablefootmark{b} \\
Gaussian mixtures     & $n$        & No  & \texttt{sklearn}\tablefootmark{b} \\
Birch                 & $n$        & No  & \texttt{sklearn}\tablefootmark{b} \\
Friend of Friends     & $n \log n$ & No  & \texttt{pyfof}\tablefootmark{c} \\
HDBSCAN               & $n \log n$ & Yes & \texttt{HDBSCAN}\tablefootmark{d} \\
\hline

\end{tabular}

\tablefoot{
\tablefoottext{a}{Runtime scalings are best case estimates and are only given with respect to number of data points $n$.}
\tablefoottext{b}{\href{https://scikit-learn.org/}{https://scikit-learn.org/}}
\tablefoottext{c}{\href{https://pypi.org/project/pyfof/}{https://pypi.org/project/pyfof/}}
\tablefoottext{d}{\href{https://pypi.org/project/hdbscan/}{https://pypi.org/project/hdbscan/}}
}

\end{table}

While many clustering algorithms exist in the literature, the complexities of \emph{Gaia} data make only a few appropriate for a large-scale unsupervised OC search. In future works, we will run on the $\approx 200$ million stars in \emph{Gaia} data brighter than $\text{G}=18$, only a small fraction of which reside in OCs. Individual fields can contain up to approximately five million stars. Hence, any clustering algorithm would need to be extremely efficient at searching through a large quantity of data to find rare objects that require a high degree of sensitivity to detect.

In later parts of this work, we compare the performance of the clustering algorithms we selected. However, to be selected for further study, the algorithms must be even remotely practical for use with \emph{Gaia} data. We set the following basic requirements on clustering algorithms for inclusion in this study. Firstly, it must be fast enough to run on the entire \emph{Gaia} dataset with a few weeks of wall time on a relatively powerful computer. Secondly, it must be able to deal with unclustered field stars (noise), as only a small fraction of stars in the Milky Way reside in OCs and the rest must be discarded. Finally, an open-source implementation must be readily available in the literature for the algorithm.

The performance of all clustering algorithms against these criteria listed in the \texttt{scikit-learn} \citep{pedregosa_scikit-learn_2011} Python library, in addition to two other common algorithms considered here, is listed in Table~\ref{tab:algorithms}. The galaxy cluster detection algorithm AMICO \citep[Adaptive Matched Identifier of Clustered Objects,][]{bellagamba_amico:_2018} was also investigated for this work, but necessary modifications to the algorithm were not made in time to adapt it for use with \emph{Gaia} data. AMICO was a top performing algorithm on mock \emph{Euclid} data \citep{euclid_collaboration_euclid_2019}, and so its application to OC detection would still be worth investigating in the future.

The first criterion disqualifies the vast majority of clustering algorithms in the literature. Practically, algorithms with runtime complexities of $\mathcal{O} ( n^2 )$ or worse (where $n$ is the number of stars) are too slow to run on large segments of the \emph{Gaia} dataset. For instance, while OPTICS \citep{ankerst_optics_1999} has seen some use in astronomy analysing smaller portions of the \emph{Gaia} dataset -- such as by \cite{ward_not_2020}, who used OPTICS to detect OB associations -- its $\mathcal{O} ( n^2 )$ runtime complexity was found to be prohibitively slow for inclusion in this work.

The second criterion favours density-based clustering algorithms such as DBSCAN \citep[Density-Based Spatial Clustering of Applications with Noise,][]{ester_density-based_1996} and HDBSCAN \citep[Hierarchical DBSCAN,][]{hutchison_hdbscan_2013}, which are the  only class of clustering algorithm that can discard points that are not in locally dense regions. These algorithms use nearest-neighbour distances to infer the local density around points, with points in low density regions discarded as field stars. However, some success has also been had in the literature with using fast algorithms to partition all data and only keep partitions that look like OCs, such as with Gaussian mixture models \citep[hereafter GMMs]{dempster_maximum_1977} by \cite{cantat-gaudin_gaia_2019}. A simple cut on proper motion dispersion can be enough to discard most non-OC partitions. The third criterion is unrestrictive, as open-source implementations exist for all algorithms that will be considered in this study.

Three algorithms were selected for further study. Firstly, DBSCAN, as mentioned previously, is a fast density-based clustering algorithm with excellent scalability, that has already proven itself in the literature in the blind searches of \cite{castro-ginard_new_2018, castro-ginard_hunting_2019, castro-ginard_hunting_2020}, recently finding hundreds of new OCs in \emph{Gaia} data. 

The second algorithm, HDBSCAN, is also density-based but improves upon DBSCAN by clustering the data hierarchically, allowing it to deal with areas of different densities better and theoretically giving it greater sensitivity. Its parameters are different to DBSCAN, and may or may not be easier to tune. It has been used by \cite{kounkel_untangling_2019} and \cite{kounkel_untangling_2020} to probe the \emph{\review{Gaia}} dataset for spatially correlated moving groups within 3 kpc, but has never been used purely to search for OCs and across all distance scales in the \emph{Gaia} dataset.

Finally, GMMs were selected for trial, an algorithm unlike DBSCAN or HDBSCAN in that it must partition all data, and partitions not containing OCs must be discarded. In principle, this could be a fast method, as GMMs have $\mathcal{O} ( n )$ runtime. A method similar to that of \cite{cantat-gaudin_gaia_2019} should be used to discard unclustered field stars with this algorithm. In addition, since it fits a model directly to the data instead of using nearest-neighbour distances, it should be less sensitive to the preprocessing or underlying shape of the data.

Some algorithms were not included in this study as their performance is clearly superseded by one of the three above. K-Means \citep{macqueen_methods_1967} is a partitioning algorithm similar to GMMs that fits a user-specified number of centroids to a dataset. Points are assigned to their nearest centroid. However, this algorithm was found to perform poorly on the \emph{Gaia} dataset, as the five dimensions of the data have intrinsically different scales. A distant cluster will have a near-negligible size in positional space, but will still form a Gaussian clump in proper motion and parallax spaces due to the dominance of \emph{\review{Gaia}} errors at these distances. Alternatively, a nearby cluster will have large sizes in position and proper motion spaces, but still a relatively small parallax dispersion as all stars are at roughly the same distance. K-Means will routinely over or under-select cluster stars without extremely careful pre-processing, as it cannot re-scale its model independently for each axis of the data. However, GMMs can, since they fit a multivariate Gaussian. As such, including K-Means in this study was unnecessary, as GMMs are effectively a generalisation of the K-Means algorithm that allows each cluster to have a covariance matrix (i.e. a different scale for each axis.) In a similar vein, while Birch \citep{zhang_birch_1996} is similar to K-Means clustering but makes a number of improvements, it also struggles to deal with clusters that have different scales in each dimension for the same reasons.

The Friend of Friends (FoF) algorithm has also been used in the literature \citep{liu_catalog_2019} and has a history of use in astronomy, especially in searches for dwarf galaxies \citep[e.g.][]{duarte_how_2014} or dark matter haloes. However, it was not included in this study as it is the same as running DBSCAN with the minimum number of points ($m_{Pts}$) parameter set to 1, since both algorithms use a global density parameter and a notion of core or border points -- or `friends' and `friends of friends' in the FoF algorithm. However, the addition of $m_{Pts}$ to DBSCAN allows it to deal with unclustered points by discarding small clusters, which is a clear advantage for \emph{Gaia} data and in line with our second criterion -- whereas users of the FoF algorithm must manually discard small clusters.

In the following sub-sections, each selected algorithm will be explained in brief detail, along with any steps necessary to determine their parameters.

%--------------------------------------------------------------------
\subsection{DBSCAN}

\begin{figure}
   \centering
   \includegraphics{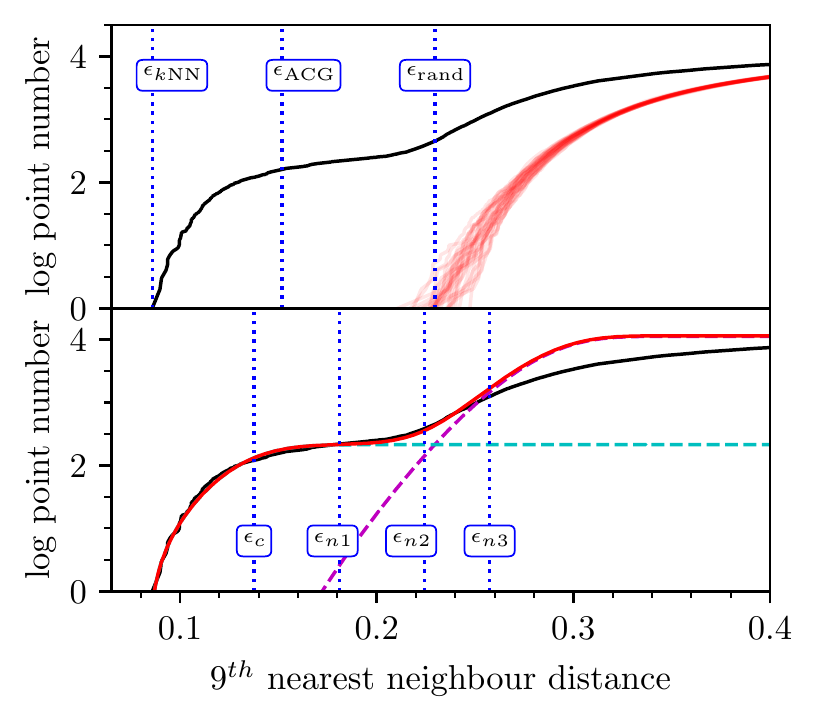}
   \caption{Nearest neighbour graphs for both methods of determining the optimum $\epsilon$ for DBSCAN. To produce this plot, which is effectively an unnormalised log cumulative density function (CDF) of nearest neighbour distances, stars are sorted based on their $k^{\text{th}}$NN distances and numbered from one to $n$. These labels as a function of $k^{\text{th}}$NN distance are then plotted to form a continuous curve. The black line on both plots is the $9^{\text{th}}$NN distances of a 2.5$^\circ$ field around the nearby OC Blanco 1. On the upper plot, $\epsilon$ estimates are determined by re-sampling the field 30 times (shown in red) to smooth out the signature of clustered stars. On the lower plot, a model of the signature of the cluster (cyan, dashed) and the field (magenta, dashed) is summed (red, solid) to approximate the curve and produce four $\epsilon$ estimates.}\label{fig:dbscan}%
\end{figure}

\subsubsection{Description of algorithm}

DBSCAN \citep{ester_density-based_1996} is one of the oldest and most widely used density-based clustering algorithms in the literature. It works by using the distances between points as a proxy for the local density of an area in a dataset, with the densest areas labelled as clusters and sparse regions labelled as unclustered background noise. Clusters are selected using two parameters. Firstly, points in a dataset are labelled as $\epsilon$-reachable if the distance between them is lower than some threshold $\epsilon$. Secondly, points are labelled as core points if they are $\epsilon$-reachable to at least $m_{Pts}$ other points, or border points if they are not core points but are $\epsilon$-reachable to a core point, where $m_{Pts}$ also includes the considered point itself. Finally, clusters are selected as density-connected groups of points that are $\epsilon$-reachable via a core point, with all other points labelled as noise.

In this way, it follows that setting $\epsilon$ to a very large value would cause all points to be labelled as one cluster, and setting $\epsilon$ to a very small value would cause no points to be labelled as cluster members. The key is to set $\epsilon$ to an appropriate value, such that separate clusters are not accidentally merged by the algorithm, and such that the algorithm is still sensitive to sparse clusters that are only marginally denser than surrounding noise points. However, this is difficult for datasets of variable density, since $\epsilon$ is a global parameter. This is a particular issue for \emph{Gaia} data, since the density of the dataset is highly variable: due to the spherical projection of \emph{Gaia} data, the density of the dataset changes with distance, since higher distances sample a larger angular volume. In addition, fields that include opaque clouds have variable densities on scales of less than 1$^\circ$, since the high levels of extinction in the clouds reduces the completeness of the \emph{Gaia} instrument. Hence, a key challenge with using DBSCAN on \emph{Gaia} data is choosing values of $\epsilon$ that are a good enough fit to the entirety of every field under study.

The $m_{Pts}$ parameter must be set high enough to restrict the core point label to only the most densely connected points, but not so high that even real clusters do not contain enough points to generate core points. In practice, $m_{Pts}$ and $\epsilon$ do not act independently, with a different choice of $\epsilon$ able to largely reproduce the same result for most values of $m_{Pts}$. As such, $m_{Pts}$ can be set to the most efficient choice. \cite{ester_density-based_1996} suggest setting $m_{Pts}$ to twice the number of dimensions of the dataset, as higher values are more computationally intensive but do not appear to include more information. For the 5D \emph{Gaia} dataset, this would imply setting $m_{Pts}=10$, which also sets a threshold on the minimum size of an OC candidate at ten stars.

By far the most computationally expensive part of the algorithm is the computation of nearest neighbour distances. This is greatly sped up by using a $k$-d tree to calculate nearest neighbour distances efficiently, which is used by the \texttt{scikit-learn} \citep{pedregosa_scikit-learn_2011} implementation of DBSCAN which is used in this work.

In the following subsections, two methods for determining $\epsilon$ for each field are presented, which are both be compared by this study.

\subsubsection{Parameter determination with the Castro-Ginard et al. (ACG) method}

\cite{castro-ginard_new_2018} have developed a method for determining $\epsilon$ for \emph{Gaia} data that exploits the random, unclustered nature of field stars to produce consistent $\epsilon$ estimates (hereafter abbreviated as the ACG method.) A brief description of how it works follows. 

Firstly, a $k^{th}$ nearest neighbour graph is computed for the dataset, where $k = m_{Pts} - 1$ (since $m_{Pts}$ includes each point itself whereas $k$ is the distance to the nearest neighbouring point.) The smallest $k^{th}$ nearest neighbour distance $\epsilon_{k\text{NN}}$ is recorded. 

Secondly, the data are randomly re-sampled according to the overall distribution of astrometric parameters in a given field. Assuming that the contribution of a cluster to this distribution is small as very few stars reside in OCs, the signature of the cluster is removed in the randomly redrawn nearest neighbour graph, allowing it to approximate the distribution of field stars in the dataset. Its minimum $k^{th}$ nearest neighbour distance $\epsilon_{\text{rand}}$ is recorded. This step can be repeated multiple times to take a more accurate mean value of $\epsilon_{\text{rand}}$. \cite{castro-ginard_new_2018} repeat this step 30 times.

Finally, the average of these two values $\epsilon_{\text{ACG}} = (\epsilon_{k\text{NN}} + \epsilon_{\text{rand}}) \, / \, 2$ is used as $\epsilon$ by DBSCAN. When a cluster is present in a field, $\epsilon_{\text{ACG}}$ roughly approximates the modal $k^{th}$ nearest neighbour value for the cluster. When no cluster is present in a field, $\epsilon_{\text{ACG}} \approx \epsilon_{k\text{NN}} \approx \epsilon_{\text{rand}}$, and no clusters will be erroneously detected by DBSCAN.

The ACG method is explained in more depth in \cite{castro-ginard_new_2018}. The top panel of Fig.~\ref{fig:dbscan} shows how random re-sampling allows $\epsilon_{\text{ACG}}$ to be calculated. The implementation of this method differs slightly from the original used by \cite{castro-ginard_new_2018}, as random re-draws in the second step are performed by randomly re-using existing parameter values for stars instead of first averaging them with kernel density estimation, as this was found to produce equivalent results while being somewhat faster.

In \cite{castro-ginard_new_2018, castro-ginard_hunting_2020}, the size of the field under study and the parameter $m_{Pts}$ are also varied across a number of different values, helping to reduce the effect of DBSCAN's global density parameter and detect OCs of different densities. Instead, we trialed varying only $\epsilon$ with the following method, as we expect this will produce similar results while being more computationally efficient: changing the size of the field requires re-calculating the array of nearest neighbour distances, whereas only varying $\epsilon$ means that the array can be cached and efficiently re-used for new parameter values. In practice, one could also vary $\epsilon$ with the ACG method by using different multiples of $\epsilon_{\text{ACG}}$ (e.g. $1.5 \cdot \epsilon_{\text{ACG}}$ or $2 \cdot \epsilon_{\text{ACG}}$).

\subsubsection{Parameter determination with a model-fitting method}

While consistent, the ACG method is slow. Since $k^{th}$ nearest neighbour determination is the most computationally expensive part of DBSCAN, repeating it 30 times to randomly estimate $\epsilon_{\text{rand}}$ increases the runtime of DBSCAN by a factor of about 30. Instead, a model-fitting method was devised in this study to perform fast approximate analyses of the $k^{th}$ nearest neighbour graph of a field.

Instead of numerically differentiating this graph to find turning points and hence an optimum value for $\epsilon$, fitting a simple, approximate model is significantly more consistent and numerically stable. A cluster can be made up of just a few dozen stars projected against tens of thousands of background stars, complicating numerical differentiation since the signal of a cluster in such a graph is small and noisy.

\cite{chandrasekhar_stochastic_1943} derived a law for the nearest neighbour distribution of a uniformly distributed set of points in 3D, which can be converted to an arbitrary dimensionality $d$ as:

% f = r_range^(dimension + k - 1) / a^dimension * exp(-(r_range/a)^dimension) - NOT NORMALISED
% a = epsilon_max / (((k - 1) / d + 1) ** (1 / d))
\begin{equation}\label{eqn:chandrasekhar_thing}
    P \left( x, \, A, \, a, \, d, \, k\right) = 
    A \frac{x^{d + k - 1}}{a^d} \exp \left[ - \left( \frac{x}{a} \right) ^d \right],
\end{equation}

\noindent
where $x$ is the $k^{th}$ nearest neighbour distance, $A$ is a normalisation constant calculated numerically, and $a$ is a constant that can be expressed in terms of the modal $k^{th}$ nearest neighbour distance $x_{max}$ as 

\begin{equation}\label{eqn:xmax}
    a = x_{max} \left( \frac{k-1}{d} + 1 \right)^{\frac{-1}{d}}.
\end{equation}

Many different density scales exist across the 5D \emph{Gaia} dataset. OCs or globular clusters are the densest regions, with spatially correlated moving groups \citep[such as those found by][]{kounkel_untangling_2019, kounkel_untangling_2020} also forming dense groups. Unclustered field stars exist across a range of different densities: fewer stars further from the \emph{Gaia} instrument are bright enough to be detected, so distant or dust-obscured regions have lower densities; while regions in the galactic thin disk (especially towards the galactic centre) have high densities. 

Ideally, this complicated structure would be captured by fitting many instances of Eqn.~\ref{eqn:chandrasekhar_thing} simultaneously, effectively integrating across all density levels to perfectly fit a $k^{th}$ nearest neighbour model to a field. However, this would be time intensive, and was found to be unnecessary, since a simple two-instance fit in log-log space could achieve good results in less than a second of runtime. A single function $P_c$ with parameters $\theta_c = \left\{ a_c, \, d_c, \, k \right\}$ was combined with a single function $P_f$ with parameters $\theta_f = \left\{ a_f, \, d_f, \, k \right\}$, where the former and the latter represent the signal of the cluster and the field respectively:

\begin{equation}
P_{total}(x, \, A_t, \, C, \, \theta_c, \, \theta_f) = A_t \left[ C \cdot P_{c}(x, \, \theta_c) + (1-C) \cdot P_{f}(x, \, \theta_f) \right]
\end{equation}

\noindent
where a single normalisation constant $A_t$ was used. $C$, a number between 0 and 1, represents the cluster fraction, corresponding to the strength of the signal of the cluster in the $k^{th}$ nearest neighbour graph relative to unclustered field stars. $k$ was set to 9, and the fit was constrained using Eqn.~\ref{eqn:xmax} such that $x_{max, c} < x_{max, f}$.

The fit was further stabilised by finding $x_{max, f}$ numerically in a histogram of $k^{th}$ nearest neighbour distances, which was then used in conjunction with Eqn.~\ref{eqn:xmax} to fix $a_f$, leaving just four free parameters: $a_c$, $d_c$, $d_f$ and $C$, with $A_t$ determined numerically at every fitting iteration. The dimensionalities $d_c$ and $d_f$ were allowed to be non-integer to give the fit access to a greater range of shapes.

An example fit is shown in Fig.~\ref{fig:dbscan}. \review{Once a field has been fit, }points of interest in \review{the curve can be used to estimate $\epsilon$. The first, }$\epsilon_c$\review{, }is the modal $k^{th}$NN distance of the cluster component of the model, \review{and physically corresponds to the most optimum $\epsilon$ value for the most prominent cluster in a given field. $\epsilon_c$ was often similar to $\epsilon_{\text{ACG}}$. 
}

\review{When multiple OCs are in a single field, sparser objects with less contrast against field stars were not detected at the $\epsilon_c$ level, and so we also took three additional values from the curve: }$\epsilon_{n1}$ is the first inflection point in the second derivative of the overall model, $\epsilon_{n2}$ is the point in the model with the highest second derivative (i.e.\ the highest rate of change of curvature) and $\epsilon_{n3}$ is the third inflection point in the second derivative of the overall model. \review{These additional points correspond to where unclustered stars become increasingly dominant in the nearest neighbour distribution of the entire field, and allow low contrast objects in a field to be detected even if the shape of the fit and the value of }$\epsilon_c$ \review{has been primarily influenced by a denser object in the field. However, there is a trade-off: these higher values of $\epsilon$ are also likely to produce more false positives. Values higher than $\epsilon_{n3}$ were briefly investigated but were found to have false positive rates that were too high to be useful.
}

%DIF >  Other methods for determining $\epsilon$ exist and could be explored in future works. For instance, in the search of \cite{liu_catalog_2019} for OCs with the FoF algorithm (which is closely related to DBSCAN), the authors set the linking length parameter $r$ (which is similar to $\epsilon$) based directly on the number of stars in a given field $N$, as $r = 0.2 / N^{1/5}$. The authors reported that such a simple criterion produced noisy results that required post-processing, although it does have the advantage of being significantly easier and faster to compute than the two methods presented above. 

%DIF >  Ultimately, different datasets in astronomy will have different requirements. Detecting OCs within the highly variable densities in the entire \emph{Gaia} dataset benefits from the more sophisticated and data-driven methods presented above, which help to reduce the number of reported false positives. On the other hand, clusters in datasets with a simpler density structure can likely be revealed just as well with more simple methods, or even just by eye. especially when the expected physical density and size of clusters is well known beforehand, which could be used to calibrate $\epsilon$.

%--------------------------------------------------------------------

%--------------------------------------------------------------------

\subsection{HDBSCAN}\label{sec:hdbscan}

\subsubsection{Description of algorithm}

\begin{figure}
   \centering
   \includegraphics{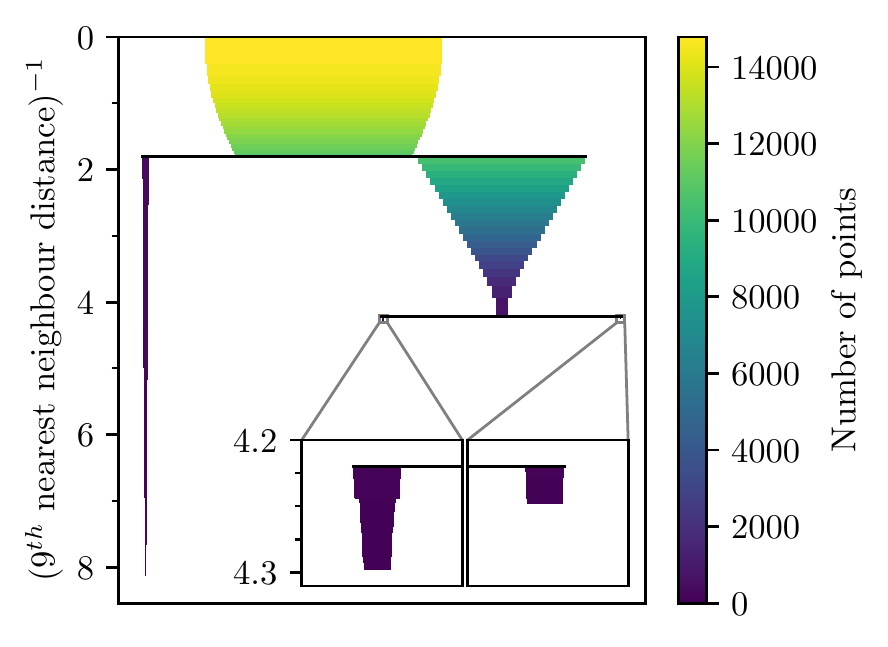}
   \caption{Condensed tree graph for HDBSCAN with $m_{clSize}=80$ applied to a $2.5^{\circ}$ field around Blanco 1, a nearby cluster without any other known OCs in the field. The colour and width of each icicle denotes the number of stars remaining in the cluster. Horizontal splits occur when clusters are no longer connected. The long icicle on the left is Blanco 1, which is an extremely clear, nearby cluster and hence splits early from field stars. On the right, the algorithm continues discarding field stars, splitting into two very short icicles at the end which are false positive clusters. The two small sub-plots in the lower right are zoomed in on the two small icicles.}\label{fig:hdbscan}%
\end{figure}

HDBSCAN \citep{hutchison_hdbscan_2013} is a more recently developed clustering algorithm \review{that attempts to improve the performance and usability of previous approaches. HDBSCAN combines the density-based approach of DBSCAN with hierarchical clustering, allowing it to deal with datasets of varying densities. Despite the extra computations, HDBSCAN does not have a significant increase in runtime compared to DBSCAN. 
}

\review{To evaluate possible clustering, nearest neighbour distances are calculated }as with DBSCAN\review{. However, HDBSCAN then }effectively considers all possible DBSCAN solutions for all possible values of $\epsilon$, \review{constructing a hierarchical tree }representation of the \review{possible clusterings of the dataset. As }with DBSCAN, \review{clusters are defined using an $m_{Pts}$ parameter to define }core and border points. \review{HDBSCAN `replaces' the }$\epsilon$ \review{parameter of DBSCAN with a }minimum cluster size $m_{clSize}$\review{, which is used to define the minimum possible size of a cluster before all points within it are instead classified as noise}. Smaller values of $m_{clSize}$ cause the hierarchical graph to be split more, as deeper, more nested solutions become valid. Larger $m_{clSize}$ values will merge small groups, negating the algorithm's sensitivity to clusters smaller than $m_{clSize}$ but while reducing the number of \review{false positive }associations of points in the dataset that are reported as clusters.

Figure~\ref{fig:hdbscan} shows a representation of the HDBSCAN hierarchical graph for clustering analysis performed on a 2.5$^\circ$ field centred on Blanco 1, with parameters $m_{clSize}=80$ and $m_{Pts}=10$. \review{Having produced a hierarchical graph representation of the dataset, clusters }can be selected \review{from it }in one of two ways. In the Excess of Mass (EoM) method, the clusters with the largest area in this plot are selected. Alternatively, in the leaf method, more fine-grained structure is revealed, as clusters at the bottom of the tree are always selected. 

HDBSCAN solves a number of issues encountered by previous approaches in the literature. For instance: whereas DBSCAN requires setting the $\epsilon$ parameter homogeneously across an entire dataset, giving it poor performance when detecting clusters of different densities, HDBSCAN's consideration of all DBSCAN solutions simultaneously gives it equal sensitivity across all density ranges of a dataset. \review{In addition, $m_{clSize}$ is a much more intuitive parameter to set than $\epsilon$ for detecting OCs, since the minimum allowable size of an OC can be decided beforehand and does not require an additional method to try and estimate it for a given dataset as with $\epsilon$.
}

However, use of HDBSCAN comes with some challenges when running on largely unclustered data - such as the \emph{Gaia} dataset, where very few stars reside in OCs. HDBSCAN is sensitive to all regions of a dataset where points appear statistically more clustered than the local background, particularly when setting the parameter $m_{clSize}$ low to ensure that HDBSCAN is sensitive to the smallest galactic OCs. In the \emph{Gaia} regime, it is unsurprising that a field of one million stars will contain many low signal to noise ratio false positive associations, where groups of 10$-$20 stars will appear more clustered than the background by statistical chance. These false positives will be reported by HDBSCAN and must be later removed to use HDBSCAN successfully at high sensitivities.

The Python implementation of HDBSCAN by \cite{mcinnes_hdbscan_2017} was used for this work, which differs from the original publication in a few small ways (such as using a $k$-d tree for nearest neighbour computation) that allow the algorithm to run faster. 

\subsubsection{Parameter tuning}

HDBSCAN parameters were straightforward to set in a number of small experiments conducted on well-characterised OCs. While the original HDBSCAN paper recommends setting $m_{Pts}$ and $m_{clSize}$ to the same value, setting $m_{Pts}=10$ was found to offer the best sensitivity and speed when running the algorithm, a decision also supported by the arguments for setting $m_{Pts}=10$ for DBSCAN.

To select candidate clusters from the hierarchical tree, the leaf selection method was almost always superior to the EoM method. OCs contain a very small number of stars ($\sim$100) compared to the fields they occupy ($\sim$100 000+), and the leaf selection method was significantly better at recovering the smallest objects (OCs) in a given field.

However, there is no perfect setting for $m_{clSize}$ \review{having investigated the effect of the parameter in a number of small experiments, for which we tested different parameter values against a representative set of OCs. An exact $m_{clSize}$ setting must be found empirically based on the properties of a dataset. While theory suggests that $m_{clSize}$ should not be smaller than the smallest size of an OC (which we define as ten stars in this work), such low values were found to produce }a large number of false positives\review{, also }sometimes erroneously splitting the largest OCs into two or more sub-clusters that miss many valid members of the cluster. Alternatively, setting it high (e.g. $m_{clSize}=80$) makes the algorithm's output significantly less noisy at the cost of missing the smallest objects. \review{Values larger than 80 had no added advantages despite further decreasing the algorithm's sensitivity to smaller OCs. }High values also sometimes select dense regions of field stars that must be removed later as they are not OCs.  They may correspond to moving groups such as those reported by \cite{kounkel_untangling_2019} and \cite{kounkel_untangling_2020}. A range of settings (10, 20, 40 and 80) will be compared in this paper.

%--------------------------------------------------------------------

\subsection{Gaussian mixture models}

\subsubsection{Description of algorithm}

GMMs differ from the other methods considered in this study in a number of ways, offering an interesting alternative viewpoint on how an entirely different and much older method performs when trying to detect OCs in a large, modern dataset. The data are assumed to be drawn from a number of Gaussian distributions, to which the algorithm fits a mixture of $m$ Gaussian components across a series of iterations. The likelihood of consecutive iterations is maximised until convergence is achieved. Covariances between dimensions allow the fitted Gaussians to have an elliptical or diagonal shape, which is important for OCs as many are elongated due to tidal effects.

Since all points must be assigned as a member of a Gaussian, GMMs do not have a natural way to deal with unclustered field stars. Instead, mixture components must be ruled out if their properties are incompatible with OCs. The means and standard deviations of mixture components in the different scaled dimensions $(\lambda, \, \phi, \, \mu_{\lambda*}, \, \mu_{\phi}, \, \varpi)$ can be quickly used as proxies for the properties of a candidate OC, with any targets wholly incompatible with an OC ruled out as groupings of field stars. We adopt a similar approach to \cite{cantat-gaudin_gaia_2019} and require that the following constraints are met on the proper motion dispersion $(\sigma_{\mu_{\lambda*}}, \, \sigma_{\mu_{\phi}})$ and the dispersion in positional space $(\sigma_{\lambda}, ,\ \sigma_{\phi})$ respectively:

% Proper motion constraint
\begin{equation}
\sqrt{\sigma_{\mu_{\lambda*}}^2+\sigma_{\mu_{\phi}}^2} \leq
  \begin{cases}
    1 \text{ mas yr}^{-1}                  & \varpi \leq 0.67 \text{ mas} \\
    1.49 \cdot \varpi \text{ mas yr}^{-1}  & \varpi >    0.67 \text{ mas} \\
  \end{cases} 
\end{equation}

% Radius constraint
\begin{equation}
\sqrt{\sigma_{\lambda}^2+\sigma_{\phi}^2} \leq
  \begin{cases}
    0.1 ^{\circ}                  & \varpi \leq 0.17 \text{ mas} \\
    \arctan(\varpi / 100)^{\circ} & \varpi >    0.17 \text{ mas} \\
  \end{cases} 
\end{equation}

\noindent
which differs from the constraints of \cite{cantat-gaudin_gaia_2019}, who only include a proper motion constraint. The addition of a latter radius constraint helps to remove clear false positives that are significantly larger than the typical size of OCs.

The implementation of GMMs freely available in \texttt{scikit-learn} \citep{pedregosa_scikit-learn_2011} was used in this study. In addition, \cite{cantat-gaudin_gaia_2019} have used UPMASK \citep{krone-martins_upmask:_2014} to verify OC candidates. However, this was deemed unnecessary for this study, as the sample of OC candidates after the application of the constraints was already relatively clean with few false positives.

\begin{figure}
   \centering
   \includegraphics{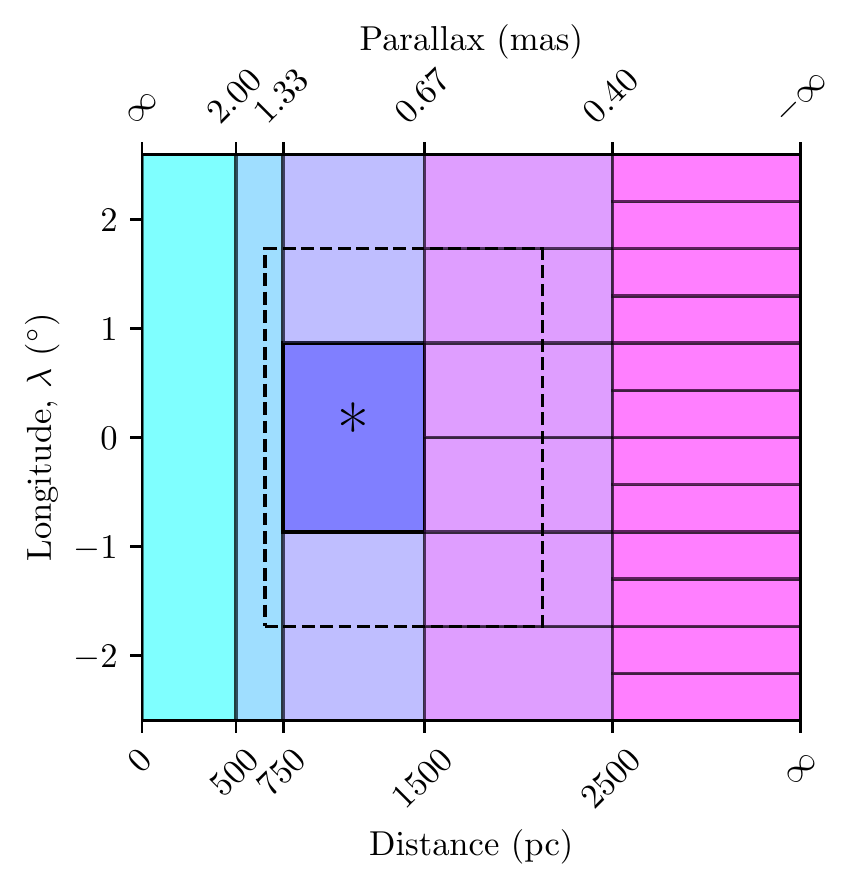}
   \caption{Schematic, top-down representation of the GMM partitioning system. Each box represents a column of sub-partitions viewed from the top. For the highlighted sub-partition also marked with an asterisk (*), the dashed width of the box shows the region in which extra stars with a parallax uncertainty of greater than 1 mas would be included. The height of the dashed box shows the extra overlap between this sub-partition and nearby other sub-partitions. Any cluster with a centroid within the dashed region but not within the main highlighted region was automatically discarded, as it will be better characterised by the neighbouring sub-partition its centroid is in.}\label{fig:gmm}%
\end{figure}

\subsubsection{Parameter tuning \& dataset sub-partitioning}

Issues were encountered when attempting to tune the number of mixtures $m$. Firstly, larger fields required linearly more mixtures to ensure that enough were available for fitting to field stars, such that $m \propto n$. Instead, it is easier to set the parameter $m_s$, the number of stars per mixture -- where $m = n / m_s$. This causes the method to be $n$ times slower, since the GMM runtime complexity also scales linearly with the number of mixtures $\mathcal{O}(nm)$, which for this choice of parameters means it is equivalent to $\mathcal{O}(n^2)$ since the number of mixtures linearly increases with the number of stars. This causes the method to fail the speed criterion (criterion one) from Sect.~\ref{sec:algorithms_criteria}, even though it was initially believed to be the fastest method under consideration.

To rectify this and ensure that GMMs can still be included in this study, a method for sub-partitioning \emph{Gaia} data chunks was devised. While \cite{cantat-gaudin_gaia_2019} used a $k$-d tree to partition fields into groups of 8000 stars, this method had no overlap between partitions, and hence may miss OCs that are split between partitions. $k$-d trees work by splitting random dimensions of a dataset along their median until each branch of the tree is small enough, a process that has no guarantee against splitting a possible OC into many different branches.

Instead, stars in a given field were divided into five segments based on parallax, where stars may be a member of any segment that they have a better than $2\sigma_{\varpi}$ agreement with. Each parallax segment was sub-divided into smaller HEALPix pixels at a specific level. Neighbouring HEALPix pixels were also selected to overlap sub-partitions between each other. The levels of the primary and overlap pixels were carefully selected to ensure that the nearest edge of every sub-partition could always fully contain an OC of 10 pc radius. In the case of the most diffuse OCs, this method could miss some stars that are far from the OC's centre, but should always be able to detect the core of all OCs. When any sub-partition in a parallax range contained fewer than $10m_s$ stars, the main HEALPix level for the parallax segment was decreased by one to increase the number of stars in the sub-partitions. This ensured that no sub-partition was impractically small for later GMM fitting. A schematic representation of this is shown in Fig.~\ref{fig:gmm}, and the values for the sub-partitions are listed in Table~\ref{tab:gmm_parameters}.

\begin{table}
\caption{Specifications of the GMM sub-partitioning scheme.}
\label{tab:gmm_parameters}
\centering
\begin{tabular}{c c c c}
\hline\hline
distance       & Max. HEALPix    & Max. sub & Optimum \\
range (pc)     & level (overlap) & partitions\tablefootmark{a} & $m_s$   \\
\hline                        
0 - 500         & None (None) & 1   & 1000 \\ 
500 - 750       & None (None) & 1   & 1000 \\
750 - 1500      & 5    (6)    & 9   & 800  \\
1500 - 2500     & 6    (7)    & 36  & 600  \\
2500 - $\infty$ & 7    (8)    & 144 & 250  \\
\hline

\end{tabular}

\tablefoot{
\tablefoottext{a}{When fewer than $10m_s$ stars were in a sub-partition, the main HEALPix level was decreased by one to make the sub-partitions a factor of four larger.}
}

\end{table}

Any OC candidate with a centroid $(\lambda, \, \phi)$ in an overlap pixel is automatically discarded, as it is assumed to be better characterised in the neighbouring sub-partition for which it would be more fully selected. The sub-partitioning scheme improved the runtime of the method by a factor of about five and the memory use by a factor of about 80. This could be improved further by reducing the pixel sizes or overlap levels, albeit at the cost of sensitivity to OCs on the boundaries between sub-partitions.

Two scenarios were tested in this study for a value of $m_s$. Firstly, $m_s$ was fixed to 800 stars per mixture component, which was found to be a good general value across the entire dataset. Secondly, $m_s$ was varied depending on the parallax range, as in Table~\ref{tab:gmm_parameters}. This was found to greatly improve the sensitivity of the method at high distances where OCs have fewer visible stars in \emph{Gaia} data and are much smaller.

Since GMMs are a method that relies on convergence, the randomly selected starting parameters of the Gaussian mixtures can affect the final result found by the method. Selecting the best result after multiple initialisations was not found to significantly improve results, so the \texttt{n\_init} parameter of the \texttt{scikit-learn} implementation was left at 1. However, the maximum number of iterations of the method, \texttt{max\_iter}, was set to 1000, to ensure that the method was always able to converge.

%--------------------------------------------------------------------

\section{Analysis}\label{sec:results}
\subsection{Evaluation criteria for clustering algorithms}

So far, we prepared \emph{Gaia} data for clustering analysis, selected three algorithms for further study, and developed techniques to optimise them for use on \emph{Gaia} data. In this section, we explain how we quantify the performance of the algorithms against each other by crossmatching to existing objects in the literature, and we present those results.

We quantify the performance of the algorithms using \review{a number of standardised statistics. For our existing literature OCs, we expect that a number of them are real, or true positives }(TP)\review{. However, literature catalogues such as MWSC have been shown to have a number of erroneous entries \citep{cantat-gaudin_clusters_2020}, which are in reality true negatives }(TN)\review{. While a perfect algorithm would report all true positive OCs, missed objects are defined as false negatives (FN). Similarly, when a putative object is erroneously reported as real, it is defined as a false }positive (FP).

\review{It is convenient to use these quantities to derive performance statistics normalised to be between 0 and 1, and so we }also derive the sensitivity, specificity and precision of the algorithms, which are defined as:

\begin{equation}
    \text{Sensitivity} = \frac{\text{TP}}{\text{TP} + \text{FN}},
\end{equation}

\begin{equation}
    \text{Specificity} = \frac{\text{TN}}{\text{TN} + \text{FP}},
\end{equation}

\begin{equation}
    \text{Precision} = \frac{\text{TP}}{\text{TP} + \text{FP}}.
\end{equation}

\noindent
\review{Effectively, the sensitivity is a measure of an algorithm's ability to detect real objects, the specificity is its ability to reject putative objects, and the precision is the fraction of reported objects that the user could expect are actually real.} It follows that a perfect algorithm would have all three quantities at 1, since FN and FP would be zero. However, this is of course unrealistic as no algorithm is likely to be perfect, and different studies may wish to prioritise different statistics over one another. For instance, a search for new OCs would wish to use an algorithm with a maximised sensitivity, such that as many new OCs as possible could be discovered -- although the precision of such a study is also of concern, so that as few false positive OCs as possible are reported. A search for existing OCs that attempts to improve the general quality of the OC census would need to maximise all three quantities, and may be especially concerned with maximising the specificity of the method used, such that as many putative literature OCs as possible can be ruled out.

We look in detail at the 100 main OCs of this study and derive sensitivity, specificity and precision statistics for all algorithms in these cases, giving the usefulness of each algorithm and parameter combination when searching for a given literature OC. Then, in the second part of our results, we derive true positive rates for all algorithms across all OCs in the fields in this study, giving supplementary information on the sensitivity of each algorithm as a function of the reported literature distance and size of the OCs.

\subsection{False positive identification}\label{sec:false_positives}

It is likely impossible to maximise both the specificity and sensitivity of any algorithm simultaneously: for all algorithms studied, increasing their sensitivity would always decrease their specificity. The detection of more true positive OCs always also resulted in more false positive OCs. We explore two techniques to reduce the number of false positives of the algorithms: firstly, by dropping all OC candidates with parameters unrealistic for an OC, and secondly, by using a density-based criterion to discard OC candidates that have a density compatible with being drawn from unclustered local field stars.

To reduce the number of false positive crossmatches -- particularly since algorithms such as HDBSCAN and DBSCAN when ran at maximum sensitivity reported over 40$\,$000 OC candidates, the majority of which are false positives -- OC candidates with mean parameters extremely incompatible with a real OC were first removed, using criteria presented in \cite{cantat-gaudin_clusters_2020}. The proper motion dispersion of OC candidates was required to satisfy

% Proper motion constraint
\begin{equation}\label{eqn:proper_motion}
\sqrt{\sigma_{\mu_{\alpha*}}^2+\sigma_{\mu_{\delta}}^2} \leq
  \begin{cases}
    1 \text{ mas yr}^{-1}                  & \varpi \leq 0.67 \text{ mas} \\
    1.49 \cdot \varpi \text{ mas yr}^{-1}  & \varpi >    0.67 \text{ mas.} \\
  \end{cases} 
\end{equation}

\noindent
The radius containing half of the members of the OC candidate was also required to satisfy $r_{50} < 20$ pc. These constraints were relatively weak, only removing a small number of clearly anomalous OC candidates that had velocity dispersions or radii that were clearly incompatible with real OCs.

\begin{figure}
   \centering
   \includegraphics{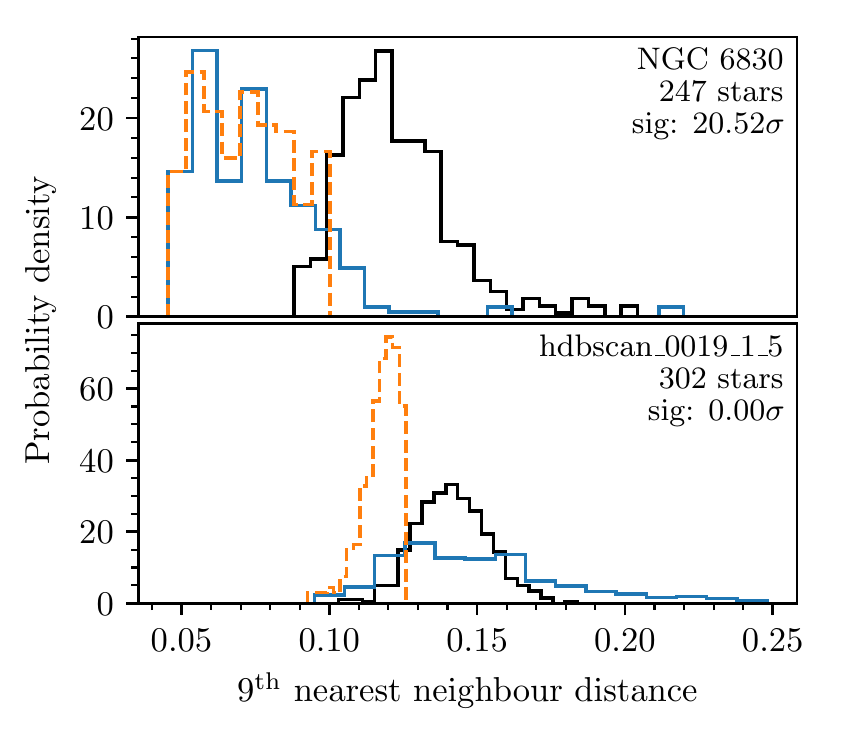}
   \caption{Two examples of NNDs used to test the significance of OC candidates. The solid black line shows the NND of nearby field stars. The blue line shows the NND of distances between cluster members. For a cluster to be significant and not simply a selection of unclustered field stars, the cluster NND must be incompatible with being drawn from the field distribution. For later illustrative purposes, the NND of a cluster member to the nearest field star is shown by the dashed orange line, although this is not used for the CST. In the upper plot, an OC candidate detected by HDBSCAN and crossmatched to the well-characterised OC NGC 6830 is shown, which has a clearly different NND to field stars with a significance of over 20$\sigma$. In the lower plot, a false positive OC detected by HDBSCAN in field 19 is shown that has a significance of 0$\sigma$.}\label{fig:nn_distances}%
\end{figure}

Secondly, a method was implemented to compare the density of OC candidates with the density of local field stars and evaluate the significance of the OC candidate, hence referred to as the cluster significance test (CST). Ninth nearest neighbour distances between stars were used as a proxy for density, as this corresponds exactly to how two of the three methods in this study performed clustering analysis (since they used $m_{Pts}=10$, i.e. $k=9$) and since this value is free of contamination from binary or multiple star systems, since they will have significantly smaller first or second nearest neighbour distances.

To calculate the density distribution of a cluster, the nearest neighbour distribution (NND) of intra-cluster distances between stars within an OC candidate was calculated. Then, in an iterative approach, a minimum of 100 and a maximum of 500 local field stars were found around the OC candidate by traversing the graph of nearest neighbours and looking for field stars with NNDs uncontaminated by proximity to the cluster, meaning that none of their $1^\text{st}$ to $9^\text{th}$ nearest neighbours were labelled as cluster members. This approach was found to generate reliable and quick approximations of the NND of local field stars.

Since a good OC candidate is a clear overdensity in the parameter space, its NND should be incompatible with being drawn from the distribution of field stars. A number of statistical tests were investigated to test this, with a Mann-Whitney U test \citep{mann_test_1947} found to be the most reliable, since it makes no assumptions about the shape of the distribution and does not require the distribution to be continuous. Significance values for each OC candidate are then derived from a one-tailed test where the alternate hypothesis is that the OC candidate has an NND incompatible with and with a lesser median than the field NND.

Requiring a CST value of at least 3$\sigma$ was found to keep the vast majority of good OCs while identifying and removing a large number of false positives for all algorithms. For instance, for DBSCAN when running with $\epsilon_{n3}$ (the algorithm and parameter combination that produced the highest number of OC candidates), the CST constraint reduced the number of reported OC candidates from 51920 to just 1111 objects.

\subsection{Crossmatches with existing catalogues}\label{sec:results_crossmatch}

Having greatly reduced the number of false positives identified by all algorithms, we crossmatched OC candidates against literature clusters to estimate the number of true positives detected by each algorithm. However, this process is non-trivial, with each catalogue reporting OCs in different ways.

A number of approaches were trialed to crossmatch OC candidates. The best approach found to crossmatch OC candidates' positions was that of \cite{liu_catalog_2019}. The tidal radius of OC candidates is estimated as the maximum distance of a member star from its mean $\alpha$ and $\delta$. The OC candidate must be within one tidal radius of the reported position in the literature, where whichever tidal radius is larger (that of the candidate or that of the literature cluster) is used. This would typically correspond to searching in a radius of no more than $0.5^\circ$.

$\mu_{\alpha*}$, $\mu_{\delta}$ and $\varpi$ for OC candidates were required to be within $5 \sigma$ (5 standard errors) of literature values. It has been shown that \emph{Gaia} DR2 has a number of small unaccounted for systematic effects, including a parallax zero-point offset $\varpi_0$ that may be magnitude-dependent \citep{lindegren_gaia_2018}. As such, even when crossmatching to other OCs detected in \emph{Gaia} data, magnitude-dependent systematic errors could cause crossmatches to fail. For instance, \cite{castro-ginard_hunting_2020} have only studied \emph{Gaia} data to $G=17$. Extra stars introduced by this study using a magnitude cut of $G=18$ will have a different mean systematic effect on derived astrometric parameters. Additionally, every clustering algorithm will report slightly different membership lists for each OC, and the differences in parameters of included or ignored members could introduce different systematic errors. In ($\alpha$, $\delta$), these effects are small, since tidal radii (often no smaller than $\sim0.1^{\circ}$) are much larger than the small systematic errors in position of the \emph{Gaia} reference frame. However, large OCs especially may have standard errors on their mean parallax or proper motion as small as $10$ $\mathrm{\mu}$as or $10$ $\mathrm{\mu}$as yr$^{-1}$, smaller than the reported \emph{Gaia} systematic errors.

To rectify missed crossmatches, small tolerances to uniform systematic errors of 50 $\mathrm{\mu}$as yr$^{-1}$ and 50 $\mathrm{\mu}$as were accounted for in crossmatching of proper motions and parallaxes respectively. These values were selected to roughly account for the scatter in parallax and proper motion offsets as a function of magnitude as reported by \cite{lindegren_gaia_2018}. This allowed a number of larger OC candidates with very small uncertainties \citep[particularly from the catalogue of][]{cantat-gaudin_clusters_2020} to be successfully crossmatched. Many of these large OC candidates were visible by eye in the \emph{Gaia} data and in the reported results of the clustering algorithms, and were being missed in the crossmatch procedure by a lack of tolerance to systematic error and due to their small uncertainties on parameters owing to their large size.

As the only non- \emph{Gaia}  catalogue, MWSC was more complicated to crossmatch against. Reported distances to OCs were converted to parallaxes. While distance measurements in MWSC do not include uncertainties, the estimated 11\% systematic uncertainty on distance measurements reported by \cite{kharchenko_global_2013} was accounted for. A parallax offset of $\varpi_0 = -0.029$ mas was applied to MWSC parallaxes, ensuring that they have the same mean systematic offset as parallaxes in the \emph{Gaia} DR2 dataset as reported by \cite{lindegren_gaia_2018}. The additional $\pm$0.8 mas yr$^{-1}$ external error in MWSC proper motions was also accounted for, which resulted in a handful of extra crossmatches to objects clearly crossmatched in other dimensions that had large offsets in their proper motions relative to \emph{Gaia} DR2.

\subsection{Results}\label{sec:analysis_results}

Finally, we present analysed results of the algorithms for discussion in three parts.

Firstly, we inspected \emph{Gaia} data manually to assign the 100 main OCs as either true positives or true negatives. An interactive data viewer was used to explore the region around the reported locations of the OCs, searching for significant overdensities within the possible crossmatch region. We also required that the detected overdensity had a colour magnitude diagram (CMD) compatible with an OC, for which we define the following criteria.

A class one OC has a clear, difficult to dispute CMD, with a realistic shape. The CMD may be somewhat broadened by differential extinction or inhomogeneities, but there should be enough stars present to make the probability of a false alarm very small. However, a class two OC is a possible OC that may be too small or too inhomogeneous for its true existence to be clear. It may be that only the brightest stars (near the turnoff point) are detected, making its shape difficult to discern as a true isochrone. There may be a small number of outlier stars incompatible with an isochrone, owing to a poor detection by the algorithm. This class signifies that more work would be needed to confirm this object as an OC. Finally, class three OCs are very unlikely to be an OC and much more compatible with random noise. Even if some stars follow an isochrone, a significant number are outliers, owing to this being a selection of unclustered, inhomogeneous stars.

After an overdensity was isolated in position, proper motion and parallax, it was required to have a class one or two CMD to confirm it as a true positive OC. We assigned 40 OCs as true positives and the remaining 60 as true negatives.

31 of the 33 OCs in the list of main OCs from MWSC that are also in the catalogue of \cite{cantat-gaudin_clusters_2020} were entered as true positives. Most of these objects were good OCs that were clearly visible at their reported location. We did not detect significant overdensities with class one or two CMDs corresponding to Patchick 75 or Auner 1, both of which are distant OCs with distances in \cite{cantat-gaudin_clusters_2020} of $\sim$7 kpc and $\sim$8 kpc respectively. These OCs are heavily polluted in the literature membership lists. If real, they are scarcely detectable in \emph{Gaia} data. Alternatively, they may simply not be real objects.

Most of the additional 67 OCs listed in MWSC do not appear detectable in \emph{Gaia} data, and may simply not be real objects. However, we did find sparse overdensities corresponding to nine objects from MWSC: ASCC 28, ASCC 100, ASCC 130, BDSB 124, Berkeley 64, DBSB 164, IRAS 06046-0603, SAI 90 and Teutsch 146. 

After reducing the number of false positives in the results of the algorithms using the techniques from Sect.~\ref{sec:false_positives}, we crossmatched their results to the main list of 100 OCs and derived performance statistics. To quantify uncertainty on derived statistics, we used the method for computing Bayesian binomial confidence intervals described in \cite{cameron_estimation_2011}, where a Beta distribution with an uninformative prior is used to estimate a confidence interval containing the true success fraction given the measured success fraction. The performance of the algorithms is listed in Table~\ref{table:classifications}. 

Five OCs from the 40 true positives are never detected by any algorithm within our constraints, which we discuss here for completeness: Berkeley 91 and Teutsch 156 \citep[objects from][]{cantat-gaudin_clusters_2020} as well as ASCC 28, BDSB 124 and Teutsch 146 from MWSC. Berkeley 91 is relatively distant ($\sim4$ kpc) OC with a polluted CMD in the catalogue of \cite{cantat-gaudin_clusters_2020}. If real, it is barely detectable in \emph{Gaia} data. Teutsch 156 appears to be detected by HDBSCAN, but only tentatively with a CST of 0.68$\sigma$. ASCC 28 should be detected, as the detected overdensity was nearby with a parallax of 0.85 mas. It may be too sparse for an algorithm to detect or may be an association mis-classified by the expert classifier. The BDSB 124 and Teutsch 146 overdensities are distant ($\varpi \approx 0.3$ mas and 0.25 mas respectively) with polluted CMDs. These objects may be difficult for algorithms to detect in \emph{Gaia} data or may simply be associations.

\review{Six OCs from MWSC listed as true negatives are reported at some point by any algorithm (five by HDBSCAN, two by DBSCAN), although only OC candidates crossmatched to FSR 0316 are detected by two different algorithms (HDBSCAN and DBSCAN). In all of these cases, the objects are sparse and relatively separated from the reported literature locations on the sky, and may be new OCs that marginally coincide with the existing locations.
}

\begin{figure*}
   \centering
   \includegraphics{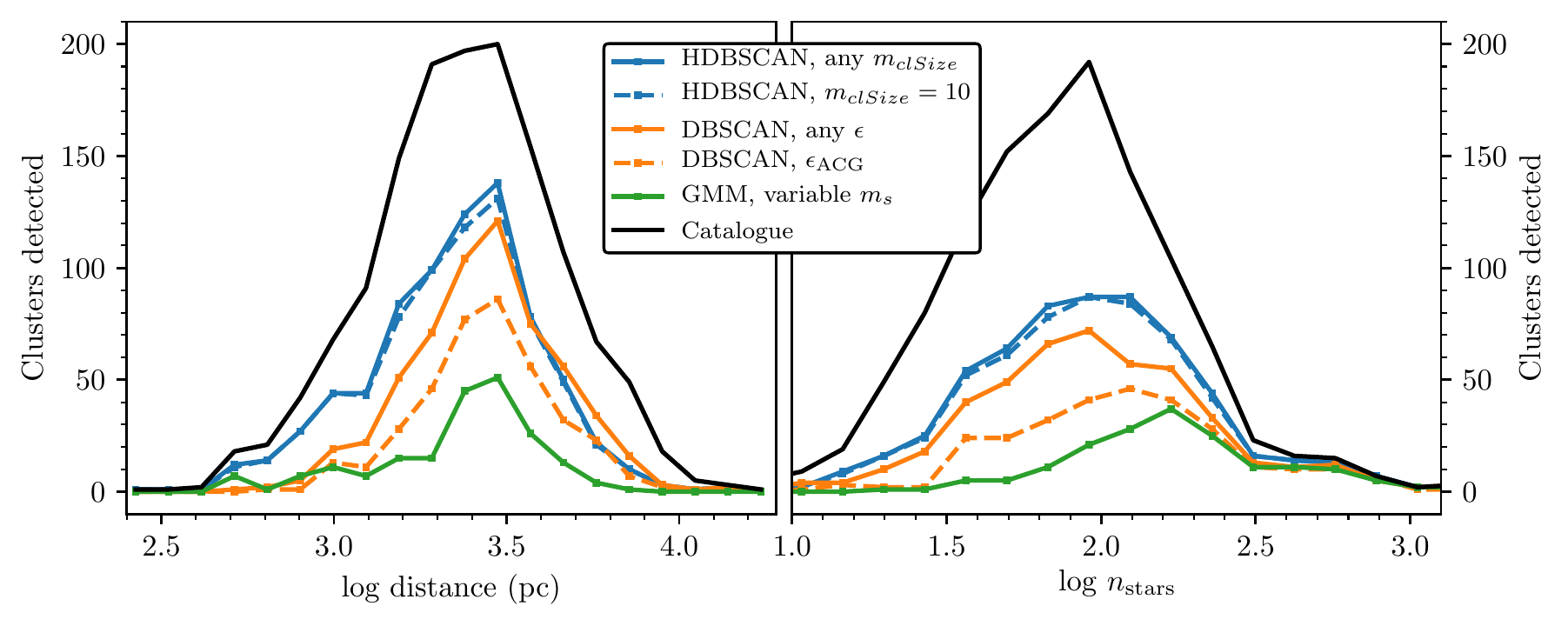}
   \caption{Distance and size dependence of detections by different algorithm and parameter combinations for all  \review{1385 }OCs in all studied fields, plotted against the reported size and distance of the OCs in the literature. OC candidates not passing the criterion in Sect.~\ref{sec:false_positives} and with a CST of less than 3$\sigma$ were discarded. HDBSCAN detects the most OCs, especially at nearby distances. GMMs only perform well at detecting well populated OCs. While individual DBSCAN results at different $\epsilon$ values do not detect especially many OCs, combining them all together nearly matches the performance of HDBSCAN -- even exceeding it slightly at large distances.}\label{fig:detections_by_distance}%
\end{figure*}

Secondly, we performed crossmatches to all \review{1385 }targets listed in the literature in the fields of this study. While there are too many OCs to conduct a precise by-hand treatment of the results, these results give a better indication of the dependence of the algorithms' sensitivities on OC features like distance, age and size. \review{Clear dependencies on distance and size are found, which are shown in Fig.~\ref{fig:detections_by_distance}. }No significant dependence on ages listed in MWSC is found for any of the algorithms, \review{although the more recent and accurate age catalogue of 
\cite{cantat-gaudin_painting_2020} did reveal a slight dependence on age that appears to result from the smaller size of older OCs. HDBSCAN is the most sensitive algorithm across all ages. When combining all $\epsilon$ runs, DBSCAN is as sensitive for well populated young OCs, but is less sensitive to older, typically smaller OCs. GMMs are the worst algorithm across all ages.
}

Finally, to compare the general usability of the algorithms, we list the runtimes and the total number of OC candidates with valid proper motion dispersions and radii reported by each algorithm in Table~\ref{tab:algorithm_performance}. Ideally, an algorithm would report a realistic number of OC candidates in as little time as possible.

We also list additional comparisons of our results with other catalogues in Appendix~\ref{app:comparison_with_cats}. \review{Full tables of detected and non-detected objects (including OC membership lists) are available in the online material only, with descriptions of their content in Appendix~\ref{app:extra_tables}.
}

% Table showing detections as a fn of class
\begin{table*}
\caption{Performance of different algorithm and parameter combinations on the 100 main OCs.}              % title of Table
\label{table:classifications}      % is used to refer this table in the text
\centering                                      % used for centering table
\begin{tabular}{l l | c c c c | c c c}          % centered columns (4 columns)
\hline\hline                        % inserts double horizontal lines

Algorithm & Parameters & TP & FP & TN & FN & Sensitivity & Specificity & Precision \\    
\hline
DBSCAN    & $\epsilon_{ACG}$, 1 repeat   & 22 $^{25.0}_{18.8}$ & 0 $^{1.8}_{0.0}$ & 60 $^{60.0}_{58.2}$ & 18 $^{21.2}_{15.0}$ & 0.55 $^{0.62}_{0.47}$ & 1.00 $^{1.00}_{0.97}$ & 1.00 $^{1.00}_{0.91}$ \rule{0pt}{0.35cm}\\[0.1cm]
$\quad$-  & $\epsilon_{ACG}$, 30 repeats & 20 $^{23.1}_{16.9}$ & 0 $^{1.8}_{0.0}$ & 60 $^{60.0}_{58.2}$ & 20 $^{23.1}_{16.9}$ & 0.50 $^{0.58}_{0.42}$ & 1.00 $^{1.00}_{0.97}$ & 1.00 $^{1.00}_{0.90}$ \\[0.1cm]
$\quad$-  & $\epsilon_{c}$               & 20 $^{23.1}_{16.9}$ & 1 $^{3.2}_{0.7}$ & 59 $^{59.3}_{56.8}$ & 20 $^{23.1}_{16.9}$ & 0.50 $^{0.58}_{0.42}$ & 0.98 $^{0.99}_{0.95}$ & 0.95 $^{0.97}_{0.84}$ \\[0.1cm]
$\quad$-  & $\epsilon_{n1}$              & 20 $^{23.1}_{16.9}$ & 0 $^{1.8}_{0.0}$ & 60 $^{60.0}_{58.2}$ & 20 $^{23.1}_{16.9}$ & 0.50 $^{0.58}_{0.42}$ & 1.00 $^{1.00}_{0.97}$ & 1.00 $^{1.00}_{0.90}$ \\[0.1cm]
$\quad$-  & $\epsilon_{n2}$              & 7 $^{10.0}_{5.2}$ & 2 $^{4.5}_{1.4}$ & 58 $^{58.6}_{55.5}$ & 33 $^{34.8}_{30.0}$ & 0.17 $^{0.25}_{0.13}$ & 0.97 $^{0.98}_{0.93}$ & 0.78 $^{0.88}_{0.54}$ \\[0.1cm]
$\quad$-  & $\epsilon_{n3}$              & 2 $^{4.4}_{1.3}$ & 0 $^{1.8}_{0.0}$ & 60 $^{60.0}_{58.2}$ & 38 $^{38.7}_{35.6}$ & 0.05 $^{0.11}_{0.03}$ & 1.00 $^{1.00}_{0.97}$ & 1.00 $^{1.00}_{0.43}$ \\[0.1cm]
$\quad$-  & $\{ \epsilon_{c}, \, \epsilon_{n1}, \, \epsilon_{n2}, \, \epsilon_{n3} \} $ & 25 $^{27.8}_{21.8}$ & 2 $^{4.5}_{1.4}$ & 58 $^{58.6}_{55.5}$ & 15 $^{18.2}_{12.2}$ & 0.62 $^{0.69}_{0.54}$ & 0.97 $^{0.98}_{0.93}$ & 0.93 $^{0.95}_{0.83}$ \\[0.1cm]

\hline

HDBSCAN   & $m_{clSize} = 80$            & 17 $^{20.2}_{14.1}$ & 3 $^{5.7}_{2.1}$ & 57 $^{57.9}_{54.3}$ & 23 $^{25.9}_{19.8}$ & 0.42 $^{0.50}_{0.35}$ & 0.95 $^{0.97}_{0.91}$ & 0.85 $^{0.91}_{0.71}$ \rule{0pt}{0.35cm}\\[0.1cm]
$\quad$-  & $m_{clSize} = 40$            & 24 $^{26.8}_{20.8}$ & 6 $^{9.2}_{4.4}$ & 54 $^{55.6}_{50.8}$ & 16 $^{19.2}_{13.2}$ & 0.60 $^{0.67}_{0.52}$ & 0.90 $^{0.93}_{0.85}$ & 0.80 $^{0.86}_{0.69}$ \\[0.1cm]
$\quad$-  & $m_{clSize} = 20$            & 31 $^{33.1}_{27.9}$ & 7 $^{10.3}_{5.2}$ & 53 $^{54.8}_{49.7}$ & 9 $^{12.1}_{6.9}$ & 0.78 $^{0.83}_{0.70}$ & 0.88 $^{0.91}_{0.83}$ & 0.82 $^{0.86}_{0.73}$ \\[0.1cm]
$\quad$-  & $m_{clSize} = 10$            & 33 $^{34.8}_{30.0}$ & 7 $^{10.3}_{5.2}$ & 53 $^{54.8}_{49.7}$ & 7 $^{10.0}_{5.2}$ & 0.82 $^{0.87}_{0.75}$ & 0.88 $^{0.91}_{0.83}$ & 0.82 $^{0.87}_{0.74}$ \\[0.1cm]

\hline

GMM       & $m_{s} = 800$                & 7 $^{10.0}_{5.2}$ & 0 $^{1.8}_{0.0}$ & 60 $^{60.0}_{58.2}$ & 33 $^{34.8}_{30.0}$ & 0.17 $^{0.25}_{0.13}$ & 1.00 $^{1.00}_{0.97}$ & 1.00 $^{1.00}_{0.75}$ \rule{0pt}{0.35cm}\\[0.1cm]
$\quad$-  & $m_{s} = $ variable          & 13 $^{16.2}_{10.4}$ & 0 $^{1.8}_{0.0}$ & 60 $^{60.0}_{58.2}$ & 27 $^{29.6}_{23.8}$ & 0.33 $^{0.41}_{0.26}$ & 1.00 $^{1.00}_{0.97}$ & 1.00 $^{1.00}_{0.85}$ \\[0.1cm]

\hline                                             %inserts single line
\end{tabular}

\tablefoot{True positive (TP), false positive (FP), true negative (TN) and false negative (FN) counts of detected clusters are given along with the sensitivity, specificity and precision. 68.3\% confidence intervals are shown for all numbers. Confidence intervals for a handful of values (e.g. measured precisions of exactly 0.0 or 1.0) were adjusted to include the measured values. This corrects for approximations in the calculation of binomial confidence intervals where the measured success probability is exactly 0 or 1. All objects that did not pass the criterion in Sect.~\ref{sec:false_positives} with a CST greater than 3$\sigma$ were discarded before crossmatching and producing this table.}

\end{table*}

% Detections vs. distance and size

%DIF >  Table with extra information about runtimes etc
\begin{table*}
\caption{\review{Extra information on the algorithms' performance.}}
\label{tab:algorithm_performance}
\centering
\begin{tabular}{l c c c c}
\hline\hline
Algorithm& Reported OC candidates\tablefootmark{a} & \review{Fraction with CST > $3\sigma$ }& \review{Total crossmatches}\tablefootmark{b} & Mean runtime (mins)\tablefootmark{c} \\
\hline

DBSCAN (ACG)    & 1518 to 1538  & \review{58.9\% to 59.6\%} & \review{382 }& \begin{tabular}{@{}l@{}}\review{1.19 (1 repeat) to }\\ 10.3 (30 repeats)\end{tabular}       \\[0.25cm]

DBSCAN (model)  & 5212 to 51920 & \review{22.4\% to 2.1\%}  & \review{593 }& 0.885 \\[0.1cm]

HDSBCAN         & 1196 to 49693 & \review{82.0\% to 5.2\%}  & \review{756 }& 2.36 \\[0.1cm]

GMM             & 314 to 2465   & \review{60.5\% to 20.5\%} & \review{213 }& \begin{tabular}{@{}l@{}}21.9 ($m_s=800$) \\ 47.0 (variable $m_s$)\end{tabular}      \\
\hline

\end{tabular}

\tablefoot{
\tablefoottext{a}{Total number of OC candidates for all fields that passed the proper motion dispersion and radius constraints from Sect.~\ref{sec:false_positives}. The range is between the minimum and maximum reported number for the least and most sensitive parameters.}
\tablefoottext{b}{Total crossmatches is given as the union of all results from all parameter sets for a given method. A total of 1385 literature OCs were crossmatched against.}
\tablefoottext{c}{Mean runtime for a single field out of the 100 in this study. All runs were conducted on the same workbench computer with a 3.1 GHz 4-core CPU.}
}

\end{table*}

%--------------------------------------------------------------------
\section{Comparison of algorithms}\label{sec:discussion}

Finally, having selected three algorithms for further study and having ran them on 100 representative fields across the galactic disk, we address the central topic of this work as to which clustering algorithm is best at detecting OCs in \emph{Gaia} data. We discuss the pros and cons of each algorithm in subsections before presenting an opinion.

\subsection{DBSCAN is effective at searching for OCs}

DBSCAN is a well proven algorithm on \emph{Gaia} data, having recently detected over 500 new OCs in \cite{castro-ginard_hunting_2020}. It performed relatively well on the 100 main OCs in this study. $\epsilon_\text{ACG}$ has particularly high specificity and precision values of $\approx1.00$ (Table~\ref{tab:algorithm_performance}), suggesting that DBSCAN can produce consistent and reliable results when not ran sensitively. However, even when greatly increasing its theoretical sensitivity (e.g. $\epsilon_{n3}$, the highest $\epsilon$ value used in this study), DBSCAN still is not able to detect all OCs present in a field.

At individual values of $\epsilon$, Fig.~\ref{fig:detections_by_distance} shows that DBSCAN is most sensitive to OCs at certain distances, with the $\epsilon_\text{ACG}$ sensitivity peaking at 3.1 kpc. This is likely due to how distant OCs are very compact in all dimensions, while nearby OCs in the sample may have radii of up to 0.5$^\circ$ or more, and are hence much sparser in the two positional dimensions and require the global density threshold $\epsilon$ to be higher for them to be completely detected. This is a key disadvantage of DBSCAN: single, global $\epsilon$ parameters rarely seem to be perfect for individual OCs, especially when the global parameter is influenced by density contributions from many different OCs in a single field. 

Manual comparison between algorithm results shows that DBSCAN often under or over-selects OCs and produces less reliable membership lists than HDBSCAN or GMMs. Over-selection is a particular issue as CMDs become polluted and the performance of OC candidates in the CST is reduced, as the nearest neighbour distribution becomes dominated by contaminating field stars. Many OC candidates detected by DBSCAN at CST values of less than 3$\sigma$ appear to correspond to real OCs, but are too polluted or too sparse to pass criterion to verify the candidate objects as real. In addition, these membership lists containing too few or too many members are of less use to other scientific applications, and would need to be followed up with another algorithm to improve their quality.

This can be partially mitigated by combining all DBSCAN results across all $\epsilon$ values, which approaches a similar degree of sensitivity to HDBSCAN, albeit still with a deficit of detections for small distances at less than 1 kpc. This result is in good agreement with what theory presented in Sect.~\ref{sec:hdbscan} suggests: that a single run of DBSCAN's global density parameter will only be sensitive to a certain size of OC at a given distance, and that running HDBSCAN is equivalent to running DBSCAN across all possible values of $\epsilon$. However, HDBSCAN is better still -- able to detect the majority of OCs in the sample in a single run at $m_{clSize}=10$. 

Combining multiple DBSCAN results is similar to the effective approach that \cite{castro-ginard_hunting_2020} use to detect over 500 new OCs, since they vary the $m_{Pts}$ parameter and the size of the field analysed by the algorithm. $\epsilon_{\text{ACG}}$ results presented here should be less sensitive than the results of \cite{castro-ginard_hunting_2020} as they are based on a single DBSCAN run at a single value of $m_{Pts}=10$ and a single size of field. However, $m_{Pts}$, $\epsilon$ and the size of the field under consideration are not entirely independent parameters, since changing $\epsilon$ has a similar effect to how changing either of the others improves DBSCAN sensitivity in \cite{castro-ginard_hunting_2020}. \review{It is not possible to quantitatively compare the sensitivity of DBSCAN methods }in this study \review{to that of \cite{castro-ginard_hunting_2020}, since they use a different cut on the }\emph{\review{Gaia}} \review{dataset ($G=17$) and autonomous CMD classification to remove false positives, which will have reduced their sensitivity to faint OCs or distant objects with high CMD contamination. They detect }688 (55.9\%) of the \review{total number of }OCs reported in \cite{cantat-gaudin_gaia_2018}, \review{although the 688 correspond to 81\% of objects in \cite{cantat-gaudin_gaia_2018} with a significant number of members brighter than $G=17$ and a well defined isochrone, which are the objects that their study would be theoretically sensitive to. The }combination of all DBSCAN runs in this study was able to detect 343 out of 537 (63.9\%) of the OCs in this study from \cite{cantat-gaudin_clusters_2020} at a CST of greater than 3$\sigma$.

Future use of DBSCAN could benefit from repeated runs while only changing $\epsilon$ opposed to the size of the field, which \review{we expect to be both as sensitive but }more computationally efficient. The most computationally expensive step (calculation of nearest neighbour distances for a given field) would only need to be performed once, with DBSCAN then evaluated quickly on the same matrix of nearest neighbour distances but simply with a wide range of different $\epsilon$ values. 

For $\epsilon$ determination for DBSCAN, both methods appear viable, although the ACG method is more numerically stable. $\epsilon_{c}$ results were largely analogous to results produced by $\epsilon_{\text{ACG}}$ -- although occasionally, on more difficult fields, the model fit would be less stable and would over-estimate the optimum value of $\epsilon$. This is clear in Table~\ref{table:classifications}, where the ACG method has a very good precision of 1.00 compared with 0.95 for $\epsilon_c$ results.

When running on large fields such as those in this study, the ACG method's random field resampling only needs to be repeated once, since no improvement is visible in crossmatch statistics between resampling 30 times versus only performing it once. The measured sensitivity of the ACG method with a single repeat is slightly better than using 30 repeats (0.55 vs. 0.50), although this difference is not statistically significant. Using only a single repeat is also an order of magnitude faster than doing 30 repeats, although the field modelling approach is faster still, being 25\% faster than the ACG method with a single repeat.

While naturally, the ACG method only produces a single $\epsilon$ estimate, it could easily produce more by using multiples of $\epsilon_{\text{ACG}}$ in the range $[1, \, 2.5]$ to approximate results between $\epsilon_c$ and $\epsilon_{n3}$ and to combine multiple different-$\epsilon$ runs to improve sensitivity. 

Overall, DBSCAN is an effective and well-proven methodology. In particular, its high precision at low sensitivities when used with the ACG method makes it an excellent choice for a limited blind search for good OC candidates. However, it was unable to detect all OCs present in a given field, and is not reliable for producing complete, minimally polluted membership lists for OCs, since $\epsilon$ is a global parameter and will inherently not be optimised for individual OCs in a given field when a field contains multiple different objects.

\subsection{HDBSCAN solves many issues encountered by DBSCAN, but is not without flaws}\label{sec:discussion_hdbscan}

Many of the issues with DBSCAN are solved with HDBSCAN. Parameter determination and setup of the method for HDBSCAN is significantly easier, since the minimum size of an OC $m_{clSize}$ is a much more intuitive choice for a parameter than one based on nearest neighbour distances for a given dataset, $\epsilon$. In terms of sensitivity, individual runs (such as $m_{clSize}=10$) are able to outperform all DBSCAN results combined, detecting the highest number of true positive OCs in the study. However, this increased sensitivity comes at a cost: HDBSCAN results generally have the worst specificity and precision scores of any algorithm in this study, with a large number of false positives and poor characterisation of true negatives (especially for $m_{clSize}=10$.) This would be even worse when not using the CST to reduce false positives: $m_{clSize}=10$ results without a CST restriction had a precision of just 0.47 and a specificity of 0.28, owing to a huge number of reported false positives. Clearly, to be used effectively, HDBSCAN must also be used with criteria to select valid clusters from its results.

This appears to happen because of how HDBSCAN autonomously decides local thresholds for if objects are or are not a cluster. Often, HDBSCAN reports OC candidates in the densest regions of the dataset. These objects are clearly not OCs, but simply features of the underlying shape of the data, since the \emph{Gaia} satellite samples a magnitude-limited spherical volume with different observed densities at different distances. This is demonstrated well by some of the existing analysis in this work. In Fig.~\ref{fig:hdbscan}, two false positive clusters were reported alongside Blanco 1 due to how HDBSCAN effectively considers all possible DBSCAN solutions, which includes erroneously reporting two small and impersistent clusterings of field stars as OC candidates. Uniform noise follows a nearest neighbour distribution given by Eqn.~\ref{eqn:chandrasekhar_thing}, which implies that field stars will have a smooth range of different nearest neighbour distances. However, when more than $m_{clSize}$ stars exist on the dense end of this curve, HDBSCAN erroneously assigns them into a cluster, even though they are simply a feature of the random nature of the unclustered stars.

This is also demonstrated by the orange curve in the lower panel of Fig.~\ref{fig:nn_distances}, where the nearest neighbour distributions of a false positive OC are plotted. The 302 stars in this false positive OC have an external (i.e. to the nearest star) density distribution that is simply a dense slice of the field star nearest neighbour distance distribution. This orange curve is analogous to what HDBSCAN and other density-based clustering algorithms use to assign stars as cluster members. However, when looking at the internal nearest neighbour distance distribution (i.e. distance to the nearest cluster member), it becomes clear that these stars are not self-consistent with being a separate, dense object, and still appear to be drawn from a nearest neighbour density distribution that is the same as that of the local field stars.

An additional issue is that increasing the sensitivity of HDBSCAN sometimes causes it to miss certain OCs. These are typically large, clearly real objects that are mistakenly split apart into multiple substructures for low $m_{clSize}$ values. To detect all OCs in a future all-sky survey, it would be necessary to run with multiple parameters and combine runs as with DBSCAN. While Fig.~\ref{fig:detections_by_distance} shows that the effect is not as significant as with DBSCAN, combining multiple runs still provides a small increase in the total number of OCs detected.

HDBSCAN detects slightly fewer OCs than the combination of all DBSCAN results for distances of greater than 4 kpc, as shown in Fig.~\ref{fig:detections_by_distance}. This appears strange at first glance, as HDBSCAN should consider all DBSCAN solutions and should in theory be able to detect all objects DBSCAN can detect. On closer inspection, it appears that this is due to HDBSCAN's approach to membership lists, since HDBSCAN includes all objects that could be cluster members but will assign them correspondingly low membership probabilities. Distant OCs are difficult to separate from field stars, as proper motions and parallaxes become decreasingly informative at large distances -- and for these objects, HDBSCAN often includes many low probability members that reduce the quality of the detection and of the CMD of the objects. 

At relatively large distances, these low probability members cause HDBSCAN to perform worse in our study. The CST does not currently consider membership probabilities, meaning that low probability members that are more likely to be members of the field would reduce the measured significance of some distant OC candidates. In the future, the CST should be modified to also include membership probabilities.

Despite some shortcomings, it appears that properly handling these (e.g.\ with \review{the CST or another }test to remove \review{false positives based on their density}) allows HDBSCAN to be used as a powerful method for OC detection. Its runtime is not significantly longer than DBSCAN (see Table~\ref{tab:algorithm_performance}), yet it is able to detect more OCs across a wider range of distances. In addition, HDBSCAN's membership lists for validated OC candidates were typically very clean, often even detecting tidal structures for OCs due to its excellent recovery of clusters across all density levels. There is room for more improvement of HDBSCAN results at distances greater than 4 kpc by optimising validation criteria to also make use of its membership probabilities.

\subsection{GMMs are inappropriate for large-scale OC blind searches in the \emph{Gaia} era}

While HDBSCAN is somewhat slower than DBSCAN, both are significantly faster than GMMs. Despite receiving the most time investment from the authors into optimising the algorithm for use on OCs, it still under-performed relative to HDBSCAN and DBSCAN by an order of magnitude in runtime. As an $\mathcal{O}(n^2)$ algorithm when used with the optimum parameters, it scales poorly to the large \emph{Gaia} dataset of many millions of stars per field in the densest regions. This is especially noticeable in the maximum single-field runtimes of GMMs. The single densest field took 20.1 hours to run for $m_s=\text{variable}$, a factor of around 40 times slower than HDBSCAN on the same field. GMMs are simply too slow for practical use with unsupervised searches through \emph{Gaia} data, and it would take many months to run on the entirety of \emph{Gaia} DR2 in its current implementation in this study without using a supercomputer.

In the test on the 100 main OCs, GMMs had the lowest maximum sensitivity of any algorithm, detecting just 33\% of the true positive OCs. However, it did perform well in the specificity and precision metrics, even without the CST. The built-in validity constraints of the GMM method on the proper motion and radial dispersion of OC candidates ensure that all reported candidates are already of high quality. This is also evident in Table~\ref{tab:algorithm_performance}, where GMMs reported a relatively small maximum number of OC candidates (2465 for the $m_s=\text{variable}$ run), although this is still not as good as the DBSCAN ACG method, for which 1538 OC candidates were reported at most -- yet the sensitivity in the study of 100 OCs was around 60\% greater, suggesting that the DBSCAN ACG method is still more efficient at producing crossmatches to existing OCs.

A further disadvantage of GMMs is their sensitivity to the number of stars per component, $m_s$: reducing this number allows the method to detect smaller OCs, but greatly increases the runtime of the algorithm, since the runtime complexity linearly scales with the number of components of the GMM. As shown in Fig.~\ref{fig:detections_by_distance}, the variable $m_s$ values used in this study are still too large to detect many smaller clusters, with the algorithm performing poorly for any objects with fewer than around 160 reported members. In addition, low values of $m_s$ begin to cause larger OCs to be erroneously split into separate objects that would require either multiple runs of the GMM algorithm at different parameters or a scheme to merge nearby clusters after a run.

More OCs may be detected by GMMs by removing outlier stars from the dataset. While this approach is not favoured by this study as it introduces biases into the running of the algorithms, cutting stars with high proper motions \citep[such as in][]{cantat-gaudin_gaia_2019} simplifies the likelihood maximisation process for the algorithm. Sometimes, the algorithm would place individual stars with extremely high proper motions into single-star clusters, since this maximises the likelihood of the overall model fit. However, this is counterproductive, as GMM components are wasted on individual stars at high proper motions and are no longer available to fit to OCs.

While this study shows that GMMs are not scalable to a large-scale blind search, it is still a useful method for deriving membership lists for the cores of OCs. GMM OC membership lists are typically very clean. When the location of an OC is known to high accuracy beforehand, GMMs can be applied quickly to a heavily cut dataset to derive a membership list. This mirrors the success of works using UPMASK to derive membership lists for existing OCs \citep[such as][]{cantat-gaudin_gaia_2018, cantat-gaudin_clusters_2020}, since UPMASK uses K-Means clustering (an algorithm closely related to GMMs) to derive OC membership lists. This approach assumes that the reported location of an OC is accurate enough to allow a dataset to be effectively cut such that an algorithm with an effective runtime complexity of $\mathcal{O}(n^2)$ can be applied in a reasonable amount of time.

%-------------------------------------------------------------------
\section{\review{New OC candidates in the galactic disk}}\label{sec:new_ocs}

\begin{table*}

% Define first header
\caption{\label{table:new_ocs_short}\reviewtwo{Mean parameters for a selection of the new OCs detected in this study.}}

\centering
\begin{tabular}{*{11}{c}}

\hline\hline

Name & $\alpha$ ($^\circ$) & $\delta$ ($^\circ$) & $l$ ($^\circ$) & $b$ ($^\circ$) & $\mu_{\alpha*}$ (mas yr$^{-1}$) & $\mu_{\delta}$ (mas yr$^{-1}$) & $\varpi$ (mas) & $r_{50}$ ($^\circ$) & $n$ & $\sigma_{\text{CST}}$ \\

\hline

PHOC 1 & 126.99 & -42.77 & 260.83 & -2.44 & -5.74 (0.03) & 4.79 (0.02) & 0.67 (0.00) & 0.11 & 32 & 8.64 \\
PHOC 2 & 280.11 & -3.75 & 28.34 & 0.73 & 0.46 (0.02) & -1.59 (0.02) & 0.36 (0.00) & 0.09 & 47 & 5.94 \\
PHOC 3 & 115.83 & -30.48 & 245.65 & -3.39 & -2.03 (0.01) & 2.35 (0.02) & 0.40 (0.01) & 0.09 & 30 & 5.72 \\
PHOC 4 & 106.79 & -7.69 & 221.57 & -0.03 & -3.80 (0.04) & 1.10 (0.02) & 0.91 (0.01) & 0.22 & 71 & 9.96 \\
PHOC 5 & 105.88 & -7.78 & 221.23 & -0.88 & -0.66 (0.01) & -1.07 (0.02) & 0.79 (0.01) & 0.13 & 39 & 6.91 \\
PHOC 6 & 280.59 & -7.23 & 25.46 & -1.29 & 0.88 (0.02) & -2.85 (0.02) & 0.38 (0.01) & 0.06 & 38 & 7.46 \\
PHOC 7 & 285.73 & 14.58 & 47.21 & 4.11 & -0.41 (0.02) & -3.15 (0.01) & 0.49 (0.01) & 0.14 & 28 & 5.92 \\
PHOC 8 & 288.83 & 14.43 & 48.47 & 1.37 & -1.69 (0.02) & -2.50 (0.02) & 0.34 (0.01) & 0.08 & 39 & 9.26 \\
PHOC 9 & 79.59 & 41.99 & 166.04 & 2.51 & 0.13 (0.02) & -0.45 (0.01) & 0.20 (0.00) & 0.07 & 43 & 6.56 \\
\multicolumn{11}{c}{$\vdots$} \\ 
PHOC 39 & 277.78 & -3.81 & 27.22 & 2.77 & 1.89 (0.05) & -8.75 (0.05) & 2.49 (0.02) & 0.48 & 139 & 15.10 \\
PHOC 40 & 287.77 & 14.27 & 47.85 & 2.21 & -1.57 (0.06) & -9.41 (0.09) & 2.99 (0.02) & 0.49 & 36 & 7.65 \\
PHOC 41 & 282.55 & 33.41 & 63.24 & 14.78 & 1.85 (0.08) & -3.84 (0.07) & 3.42 (0.02) & 0.37 & 63 & 9.42 \\

\hline

\end{tabular}

\tablefoot{\reviewtwo{Standard errors for mean proper motions and parallaxes are shown in the brackets. The full version of this table (including extra columns) is available in the online material only, following the format of Table~\ref{app:tab:cluster_lists} except with column 26 omitted.}}

\end{table*}

\subsection{\review{Methodology}}

\review{During the preparation of this work, we discovered that many of the algorithms' reported OC candidates did not crossmatch to literature targets and appeared to be distinct, new objects. We investigated this further to see if any of the objects are genuine new OC candidates.}

\review{Firstly, we made conservative cuts on our reported OC candidates to select only high-quality objects. All objects failing the criteria from Sect.~\ref{sec:false_positives} or with a CST of less than 5$\sigma$ were discarded, meaning that our sample of candidates only represents definitive astrometric overdensities. In addition, any objects with a centre closer than 1.5 estimated tidal radii to the edge of the field they were detected in were discarded, removing any objects that could have a remote possibility of issues due to edge effects.}

\review{Secondly, we performed extra crossmatching to the catalogues of \cite{dias_new_2002}, \cite{bica_multi-band_2018}, \cite{sim_207_2019}, \cite{ferreira_discovery_2020} and \cite{qin_discovery_2020}. To the best of our knowledge, they in addition to the four catalogues from earlier in this work include all reported literature OCs from at least the past two decades. }

\review{After the crossmatching and the cuts, all algorithms still appeared to detect new OCs, but the most were found by HDBSCAN. At the high CST threshold of $5 \sigma$, any objects found by DBSCAN or GMMs were almost always also found by HDBSCAN, and so for simplicity we only looked at the results of HDBSCAN with $m_{clSize}=20$, since merging the results of different algorithm and parameter combinations would be non-trivial and is beyond the scope of this work.}

\review{This produced a list of 102 tentative objects based on astrometry and crossmatching alone. A small fraction of these had CMDs that were clearly random selections of unassociated stars that followed no clear isochrone, although many others were borderline objects with poor quality CMDs. We manually selected only objects with good or relatively good quality CMDs, leaving a list of 38 new OCs. While this study was not optimised to find nearby objects, we noticed that some of the 76 objects closer than 1 kpc that were discarded because of edge effects could be real, new OCs. We investigated the 12 most promising objects by downloading new regions of }\emph{\review{Gaia}} \review{data around them and re-running HDBSCAN, finding that three of these objects are of a good quality and bringing our total to 41 new objects.}

\reviewtwo{We name the objects with the acronym PHOC (Preliminary HDBSCAN Open Cluster) as we expect to characterise these objects further in future works.} \review{\reviewtwo{Mean parameters for a selection of these objects are shown in Table~\ref{table:new_ocs_short}, with a full list of mean parameters} and members for these new objects included in the online material. \reviewtwo{Extra} descriptions of the contents of these tables \reviewtwo{are included} in Appendix~\ref{app:extra_tables}.} \reviewtwo{In addition, plots of all new objects are included in Appendix~\ref{app:new_oc_plots}.}

\subsection{\review{Comments on the new OC candidates}}

\review{We present brief remarks on some of the newly reported OC candidates. }

\review{Comparing our list of new OC candidates with the DBSCAN blind search of \cite{castro-ginard_hunting_2020} reveals patterns similar to those shown earlier in this work in Fig.~\ref{fig:detections_by_distance}. Whereas the 209 OC candidates from their work have a median distance of 2650~pc with a highest individual distance of 8400~pc, our 41 objects detected by HDBSCAN have a closer median distance of 1940~pc with a maximum distance of 4400~pc. These results are in agreement with our earlier finding that HDBSCAN is more sensitive to nearby OCs whereas DBSCAN is more sensitive to more distant OCs. However, as discussed in Sect.~\ref{sec:discussion_hdbscan}, this may simply be an artefact of our study as our CST as-implemented gives higher scores to more nearby clusters. Our future works will report more tentative candidates with lower CST scores and may be able to achieve similar sensitivities as DBSCAN with HDBSCAN.
}

\review{Our three very nearby candidates within 500 pc (\reviewtwo{PHOC} 39, \reviewtwo{PHOC} 40, and \reviewtwo{PHOC} 41) are some of the most scientifically interesting. If real, these objects demonstrate that new clusters are yet to be found even at close distances. We estimated approximate distances to these objects as the inverse parallax after correcting for the }\emph{\review{Gaia}} \review{zero-point offset \citep{lindegren_gaia_2018}, although our future works will use a more sophisticated inference-based approach. All three objects are within the galactic disk.}

\review{\reviewtwo{PHOC} 39 has an estimated distance of 396~pc, 139 member stars and a CST score of $15.1 \sigma$. While it has a broad CMD as reported, plotting only member stars with a membership probability of at least 80\% gives a much cleaner and less broadened CMD. \reviewtwo{PHOC} 40 and \reviewtwo{PHOC} 41 are more compact, composed of 36 and 63 stars respectively with CSTs of $7.7 \sigma$ and $9.4 \sigma$ and at estimated distances of just 331~pc and 290~pc. Both objects have good quality CMDs. All three OC candidates would be excellent candidates for further study (especially with spectroscopy) thanks to their proximity.}

\review{The other 38 candidates have a mean size of 49 stars, with the largest having 118 stars and the smallest 29.  Additionally, the objects had a mean CST of $7.9 \sigma$ -- many of the new OC candidates are well above our $5 \sigma$ CST threshold and represent clear astrometric overdensities.
}

%DIF > -------------------------------------------------------------------
\section{Conclusions and future prospects}\label{sec:conclusion}

In this work, we created a preprocessing pipeline for future searches of the \emph{Gaia} dataset for OCs. We selected three viable clustering algorithms from the literature and developed new methodologies to apply them effectively to the large-scale \emph{Gaia} dataset. We compare the three algorithms side-by-side on \emph{Gaia} data for the first time. We find that GMMs are an inefficient algorithm inappropriate for large-scale blind searches of the \emph{Gaia} dataset, although they are relatively effective at producing accurate membership lists of known OCs. DBSCAN is found to be feasible and successful for finding OCs, but still struggles to detect certain objects since it operates with a single global density parameter that is rarely optimal across the variable densities of \emph{Gaia} data. In particular, when DBSCAN is used with the method of \cite{castro-ginard_new_2018} for $\epsilon$ determination, we find that it has very good precision and specificity, producing only very small numbers of false positives -- although the ACG method is only sensitive to $\approx$50\% of the 40 true positive OCs in our main sample of 100. HDBSCAN is found to solve many of the issues encountered by DBSCAN and was the most sensitive algorithm of the three, although it also produces many false positives that need to be mitigated with additional post-processing. We will use HDBSCAN in future work to conduct a large-scale blind search for OCs. We expect that HDBSCAN's improved sensitivity over other methods trialed to date will reveal many more new OCs.

In addition, we detect a number of literature OCs that have previously gone undetected in \emph{Gaia} data. We expect that many more literature objects from the MWSC catalogue remain to be detected in \emph{Gaia} in future works and data releases, although the majority appear to be associations or simply undetectable in  \emph{Gaia}. We found that a handful of OCs from \cite{cantat-gaudin_clusters_2020} may be associations -- either due to being undetectable by any of the approaches we tried or due to having very poor CMDs. We hope that future work expanding our analysis to the entire \emph{Gaia} dataset will contribute further to improving the quality and completeness of the OC catalogue of the Milky Way.

\review{Finally, we searched our existing results for new objects and produced a list of 41 good quality new OC candidates, the nearest of which is at an estimated distance of just 290~pc. While many authors have performed searches for new OCs in }\emph{\review{Gaia}} \review{data, our comparison of algorithms suggests that existing surveys have gaps in their sensitivity and that many new objects are yet to be detected. Our tentative new detections demonstrate this, suggesting that the OC census is still incomplete within 2~kpc to an unknown extent. Future searches with new and improved methodologies will be essential to increase the completeness of the local OC census.
}

We plan to develop improved processes and statistical quantifiers of the strength of all OC candidate detections, including developing supervised machine learning techniques to classify OC candidate CMDs\review{, owing to their success in other works such as \cite{castro-ginard_hunting_2020}}. As methods for improved distance determination with parallaxes develop further \review{\citep[e.g. StarHorse,][]{anders_photo-astrometric_2019}}, we hope to include these in our work to increase the signal to noise ratio of OCs in the \emph{Gaia} dataset and provide cleaner membership lists. 

Data from the \emph{Gaia} satellite is overhauling our understanding of the Milky Way's structure. By continuously developing, comparing and improving our methodologies, astronomers can maximise the productivity of \emph{Gaia} data and improve our understanding of the galaxy.

\begin{acknowledgements}
\review{We thank the anonymous reviewer for their feedback which helped to improve the clarity and impact of this paper. }E.L.H. and S.R. gratefully acknowledge funding by the Deutsche Forschungsgemeinschaft (DFG, German Research Foundation) -- Project-ID 138713538 -- SFB 881 (``The Milky Way System'', subproject B5). This work has made use of data from the European Space Agency (ESA) mission {\it Gaia} (\url{https://www.cosmos.esa.int/gaia}), processed by the {\it Gaia} Data Processing and Analysis Consortium (DPAC, \url{https://www.cosmos.esa.int/web/gaia/dpac/consortium}). Funding for the DPAC has been provided by national institutions, in particular the institutions participating in the {\it Gaia} Multilateral Agreement. This research has made use of NASA’s Astrophysics Data System. This research also made use of the SIMBAD database, operated at CDS, Strasbourg, France \citep{wenger_m_simbad_2000}.

This work would not have been possible without the ready availability of many open source software projects. Not cited in the main body of text, this work made use of Python 3 \citep{van_rossum_python_2009}, NumPy \citep{oliphant_guide_2006}, SciPy \citep{virtanen_scipy_2020}, IPython \citep{perez_ipython_2007}, Jupyter \citep{kluyver_jupyter_2016}, Matplotlib \citep{hunter_matplotlib_2007}, pandas \citep{mckinney_data_2010}, Astropy \citep{robitaille_astropy_2013, astropy_collaboration_astropy_2018}, and healpy \citep{zonca_healpy_2019}.

\end{acknowledgements}


\bibliographystyle{aa} % style aa.bst
\bibliography{open_clusters_paper_1} % your references Yourfile.bib

%-------------------------------------------------------------------
\begin{appendix}

%-------------------------------------------------------------------
\section{ADQL query used to download data}\label{app:adql}

 \emph{Gaia}  DR2 data for this work was downloaded with the following ADQL query. \texttt{\{start\_number\}} should be replaced with the first possible \texttt{source\_id} of the desired pixel using Eqn.~\ref{eqn:healpix}. \texttt{\{end\_number\}} should be replaced with the first possible \texttt{source\_id} of the next integer pixel.

\begin{verbatim}
SELECT
-- Gaia astrometry
g.source_id, g.l, g.b, 
g.ra, g.ra_error, g.dec, g.dec_error, 
g.parallax, g.parallax_error, 
g.parallax_over_error, 
g.pmra, g.pmra_error, g.pmdec, g.pmdec_error, 
g.astrometric_params_solved, 

-- Gaia photometry
g.phot_g_mean_mag, g.phot_g_mean_flux, 
g.phot_g_mean_flux_error, 
g.phot_bp_mean_mag, g.phot_bp_mean_flux, 
g.phot_bp_mean_flux_error, 
g.phot_rp_mean_mag, g.phot_rp_mean_flux, 
g.phot_rp_mean_flux_error, 
g.phot_bp_rp_excess_factor, 

-- Calculate HEALPix level 5 index
GAIA_HEALPIX_INDEX(5, g.source_id) 
  AS gaia_healpix_5, 
               
-- RUWE statistics
r.ruwe, 
               
-- CBJ+2018 distances
d.r_est, d.r_lo, d.r_hi, 
d.r_len, d.result_flag 
               
-- Inner join the tables
FROM gaiadr2.gaia_source AS g 
INNER JOIN 
  gaiadr2.ruwe 
  AS r 
  ON g.source_id = r.source_id 
INNER JOIN 
  external.gaiadr2_geometric_distance 
  AS d 
  ON g.source_id = d.source_id 
           
-- Select only valid points
WHERE g.source_id >= {start_number} 
AND g.source_id < {end_number} 
AND g.astrometric_params_solved=31 
AND g.phot_bp_mean_mag IS NOT NULL 
AND g.phot_rp_mean_mag IS NOT NULL
\end{verbatim}

%-------------------------------------------------------------------
\section{Comparison with other OC catalogues}\label{app:comparison_with_cats}

We present brief comparisons with the results of other OC catalogues, in lieu of best practices proposed in \cite{cantat-gaudin_clusters_2020} and as a part of efforts towards generally improving the quality of the OC census, reporting on both positive and negative detections. In future works, we hope to expand comparisons such as this across the entire OC census, offering another viewpoint on the existence of many literature OCs.

\subsection{\cite{cantat-gaudin_clusters_2020}}

Of the 537 objects listed in \cite{cantat-gaudin_clusters_2020} and in the fields in this study, we are able to detect 86.4\% of them with at least one algorithm or parameter combination, many of which are clear overdensities with well-resolved parameters.

We single out Auner 1, Berkeley 91 and Patchick 75 from Sect.~\ref{sec:analysis_results} as objects that should be detectable but are not found by any algorithm. In addition, FSR 1460 and FSR 1509 are also undetected. If real, these objects are distant and difficult to detect in \emph{Gaia} data, although these objects also have heavily polluted CMDs in the membership lists of \cite{cantat-gaudin_clusters_2020} and hence may simply be associations. Future \emph{Gaia} data releases with better astrometric precision will shed more light on the status of these edge-case objects.

\subsection{MWSC}

We concur with the results of \cite{cantat-gaudin_gaia_2018} and \cite{cantat-gaudin_clusters_2020} that a majority of the objects in MWSC are undetectable in \emph{Gaia} data. Some of these objects may simply not be visible in \emph{Gaia} data due to reddening or large distances, although many are also likely to not be real. Future studies will have to quantify this for all OCs on a case-by-case basis. Of our 100 main OCs that were randomly selected from the MWSC catalogue, we detected OCs corresponding to 35 of them, suggesting that $\approx$35\% of the total MWSC catalogue is visible in \emph{Gaia} data. 

However, our results show that a number of MWSC objects appear to have been missed by works such as \cite{cantat-gaudin_clusters_2020}. In our larger crossmatching effort, we recovered candidates corresponding to 193 of the 607 objects listed in MWSC (31.8\%) but that are undetected in \cite{cantat-gaudin_clusters_2020}. Some of these objects may be new OCs that happen to have similar parameters to old objects, although some others are new detections of MWSC OCs in \emph{Gaia} data.

The best examples of re-detected OCs were Collinder 347, FSR 0124, FSR 0270 and FSR 1406, which were were clearly crossmatched and are clearly visible by eye in \emph{Gaia} data. In addition, Collinder 347 has also been well detected by \cite{piatti_extended_2019} in \emph{Gaia} DR2 data and recently by \cite{claria_ccd_2019} in visual spectrum photometric data. The sparse OCs Sgr OB6, Sgr OB7 and ASCC 100 were also detected, the latter of which has few members but is nearby with a parallax of 2.75 mas, suggesting that some OCs are yet to be recovered in \emph{Gaia} data even at small distances. In all seven cases, the crossmatched objects were clearly compatible in positional and distance space with MWSC values. They are also compatible in proper motion space, although at large distances the PPMXL proper motions in MWSC provide very little constraint.

While the catalogue of \cite{cantat-gaudin_clusters_2020} is the most complete and homogeneous OC catalogue to date, it still appears to lack some OCs from the literature and contains a handful of OCs that are somewhat putative. Ongoing comparisons with the results of multiple different clustering algorithms and methodologies will help to confirm, question or deny the existence of more OCs in the literature.

\subsection{\cite{castro-ginard_hunting_2020} and \cite{liu_catalog_2019}}

\cite{castro-ginard_hunting_2020} and \cite{liu_catalog_2019} have recently reported a combined total of over 600 new OCs in \emph{Gaia} DR2 data respectively. Of the \review{209 }objects from \cite{castro-ginard_hunting_2020} in the fields in this study, we detected OCs compatible with \review{135 }of them (\review{64.6}\%), representing a \review{sizable }fraction of their catalogue of new OCs that has been detected independently in \emph{Gaia} data for the first time. We note that the undetected OC UBC 638 is very close to UBC 637 (which is detected) - their reported centres are within 0.05$^{\circ}$ of one another, their proper motions 0.07 mas yr$^{-1}$ and parallaxes to within 0.1 mas, so they may be the same object.

We are able to detect OCs compatible with 24 of the 32 OCs from the catalogue of \cite{liu_catalog_2019} that are included in this study. The reasons for non-detections of OCs from both of these works remain unclear, and would need to be investigated in a future study.

%-------------------------------------------------------------------
\section{\review{Tables of detected clusters and members}}\label{app:extra_tables}

\review{Four supplementary tables are available in online-only material at the CDS}\footnote{\href{https://vizier.u-strasbg.fr/}{https://vizier.u-strasbg.fr/}}\review{. For literature clusters, all detections by all algorithms are listed following the same format as Table~\ref{app:tab:cluster_lists}.1. Any one cluster may have up to 12 different entries from detections by different algorithm and parameter combinations. When no detections were made of a literature cluster, a single blank row is given with only columns one and 26 filled. The 41 new objects have their mean parameters listed in a separate table following the format of Table~\ref{app:tab:cluster_lists}.1 except with column 26 omitted.} \review{For both literature and new OCs, members are listed in tables following the format of Table~\ref{app:tab:cluster_members}.2.}

% Cluster lists table
\begin{table}\label{app:tab:cluster_lists}
\caption{\review{Description of the tables of detected OCs.}}
\centering
\begin{tabular}{c c c l}
\hline\hline
\review{Col. }& \review{Label }& \review{Unit }& \review{Description }\\
\hline                        
\review{1     }& \review{Name                  }& \review{--  }& \review{Designation }\\
\review{2     }& \review{Internal ID           }& \review{--  }& \review{Internal designation }\\
\review{3     }& \review{Algorithm             }& \review{--  }& \review{Algorithm for detection }\\
\review{4     }& \review{Parameters            }& \review{--  }& \review{Algorithm parameters }\\
\review{5-7}\tablefootmark{a}   & \review{$\alpha$ }& \review{deg }& \review{Right ascension }\\
\review{8-10}\tablefootmark{a}  & \review{$\delta$ }& \review{deg }& \review{Declination }\\
\review{11    }& \review{$l$                   }& \review{deg }& \review{Galactic longitude }\\
\review{12    }& \review{$b$                   }& \review{deg }& \review{Galactic latitude }\\
\review{13-15}\tablefootmark{a} & \review{$\mu_{\alpha^*}$ }& \review{mas yr$^{-1}$ }& \review{Prop. motion in $\alpha \cdot \cos \delta$ }\\
\review{16-18}\tablefootmark{a} & \review{$\mu_\delta$ }& \review{mas yr$^{-1}$ }& \review{Prop. motion in $\delta$ }\\
\review{19-21}\tablefootmark{a} & \review{$\varpi$ }& \review{mas }& \review{Parallax }\\
\review{22    }& \review{$r_{50}$              }& \review{deg }& \review{Radius containing  }\\
      &                       &     & \review{50\% of members }\\
\review{23    }& \review{$r_t$                 }& \review{deg }& \review{Estimated tidal radius}\tablefootmark{b} \\
\review{24    }& \review{$n$                   }& \review{--  }& \review{Number of members }\\
\review{25    }& \review{$\sigma_{\text{CST}}$ }& \review{--  }& \review{CST score }\\
\review{26    }& \review{Source                }& \review{--  }& \review{Source catalogue }\\
\hline

\end{tabular}

\tablefoot{
\tablefoottext{a}{Where marked, three columns are provided: the mean value, standard deviation $\sigma$, and standard error $\sigma \, / \sqrt{n}$.}
\tablefoottext{b}{Estimated using the maximum distance between the centre of the cluster and an identified member star.}
}

\end{table}

% Cluster members table
\begin{table}\label{app:tab:cluster_members}
\caption{\review{Description of the membership tables for detected OCs.}}
\centering
\begin{tabular}{c c c l}
\hline\hline
\review{Col. }& \review{Label }& \review{Unit }& \review{Description }\\
\hline
\review{1     }& \review{Name                  }& \review{--  }& \review{Designation }\\
\review{2     }& \review{Internal ID           }& \review{--  }& \review{Internal designation }\\
\review{3     }& \review{Algorithm             }& \review{--  }& \review{Algorithm for detection }\\
\review{4     }& \review{Parameters            }& \review{--  }& \review{Algorithm parameters }\\
\review{5     }& \review{Source ID             }& \review{--  }& \emph{\review{Gaia}} \review{DR2 source ID }\\
\review{6-7}\tablefootmark{a} & \review{$\alpha$ }& \review{deg }& \review{Right ascension }\\
\review{8-9}\tablefootmark{a} & \review{$\delta$ }& \review{deg }& \review{Declination }\\
\review{10    }& \review{$l$                   }& \review{deg }& \review{Galactic longitude }\\
\review{11    }& \review{$b$                   }& \review{deg }& \review{Galactic latitude }\\
\review{12-13}\tablefootmark{a} & \review{$\mu_{\alpha^*}$ }& \review{mas yr$^{-1}$ }& \review{Prop. motion in $\alpha \cdot \cos \delta$ }\\
\review{14-15}\tablefootmark{a} & \review{$\mu_\delta$ }& \review{mas yr$^{-1}$ }& \review{Prop. motion in $\delta$ }\\
\review{16-17}\tablefootmark{a} & \review{$\varpi$ }& \review{mas }& \review{Parallax }\\
\review{18    }& \review{Gmag                  }& \review{mag }& \review{G-band magnitude  }\\
\review{19    }& \review{BPmag                 }& \review{mag }& \review{BP-band magnitude }\\
\review{20    }& \review{RPmag                 }& \review{mag }& \review{RP-band magnitude }\\
\review{21-22}\tablefootmark{a}& \review{G flux}& \review{$e^{-1}$ s$^{-1 }$}& \review{G-band flux}\\
\review{23-24}\tablefootmark{a}& \review{BP flux}& \review{$e^{-1}$ s$^{-1 }$}& \review{BP-band flux}\\
\review{25-26}\tablefootmark{a}& \review{RP flux}& \review{$e^{-1}$ s$^{-1 }$}& \review{RP-band flux}\\
\review{27}\tablefootmark{b}& \review{$p$}& \review{--  }& \review{Membership probability}\\
\hline

\end{tabular}

\tablefoot{
\tablefoottext{a}{Where marked, two columns are provided: the mean value and the standard error.}
\tablefoottext{b}{Always equal to one for DBSCAN as it does not produce membership probabilities for individual stars.}
}

\end{table}

%-------------------------------------------------------------------
\section{\reviewtwo{Plots of newly detected OCs}}\label{app:new_oc_plots}

\begin{figure*}[ht]
   \centering
   \includegraphics{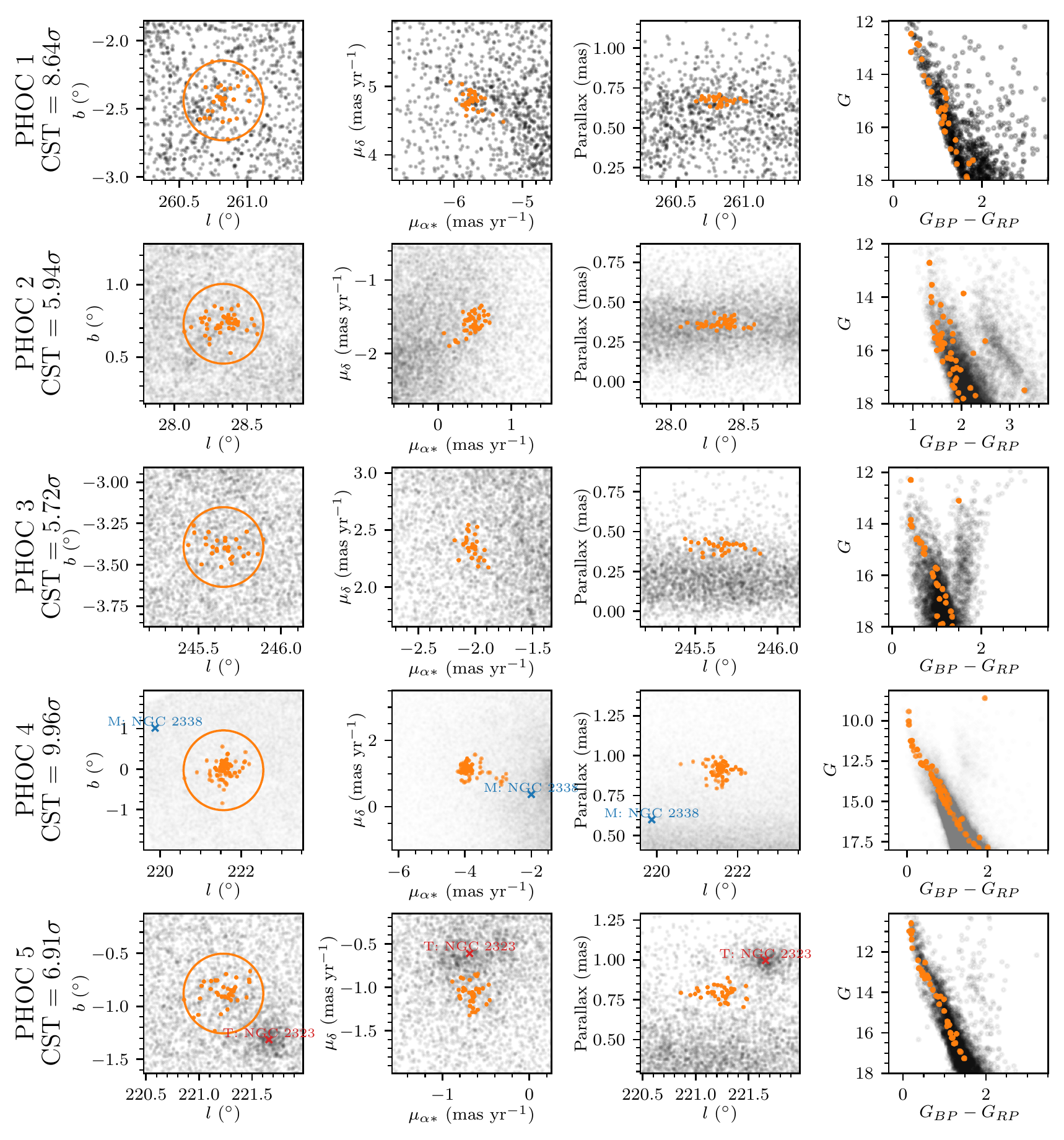}
   \caption{\reviewtwo{Astrometric and photometric plots of the first five new OCs from Sect.~\ref{sec:new_ocs}. Identified member stars are shown in orange, with background stars in black. Only members with a membership probability of greater than 50\% are plotted. The estimated tidal radius for the OCs is depicted with a circle in the $l$ vs. $b$ plots in the first column. CST scores for each object are shown with its name on the left. Nearby OCs from literature catalogues are marked when visible. T (in red text) denotes sources from \cite{cantat-gaudin_clusters_2020}, while M (blue) and S (purple) denote sources from MWSC and \cite{sim_207_2019} respectively that were not detected by \cite{cantat-gaudin_clusters_2020}. A (brown) denotes new OCs detected recently by \cite{castro-ginard_hunting_2020}.   }}\label{fig:new_ocs_0}%
\end{figure*}

\begin{figure*}[ht]
   \centering
   \includegraphics{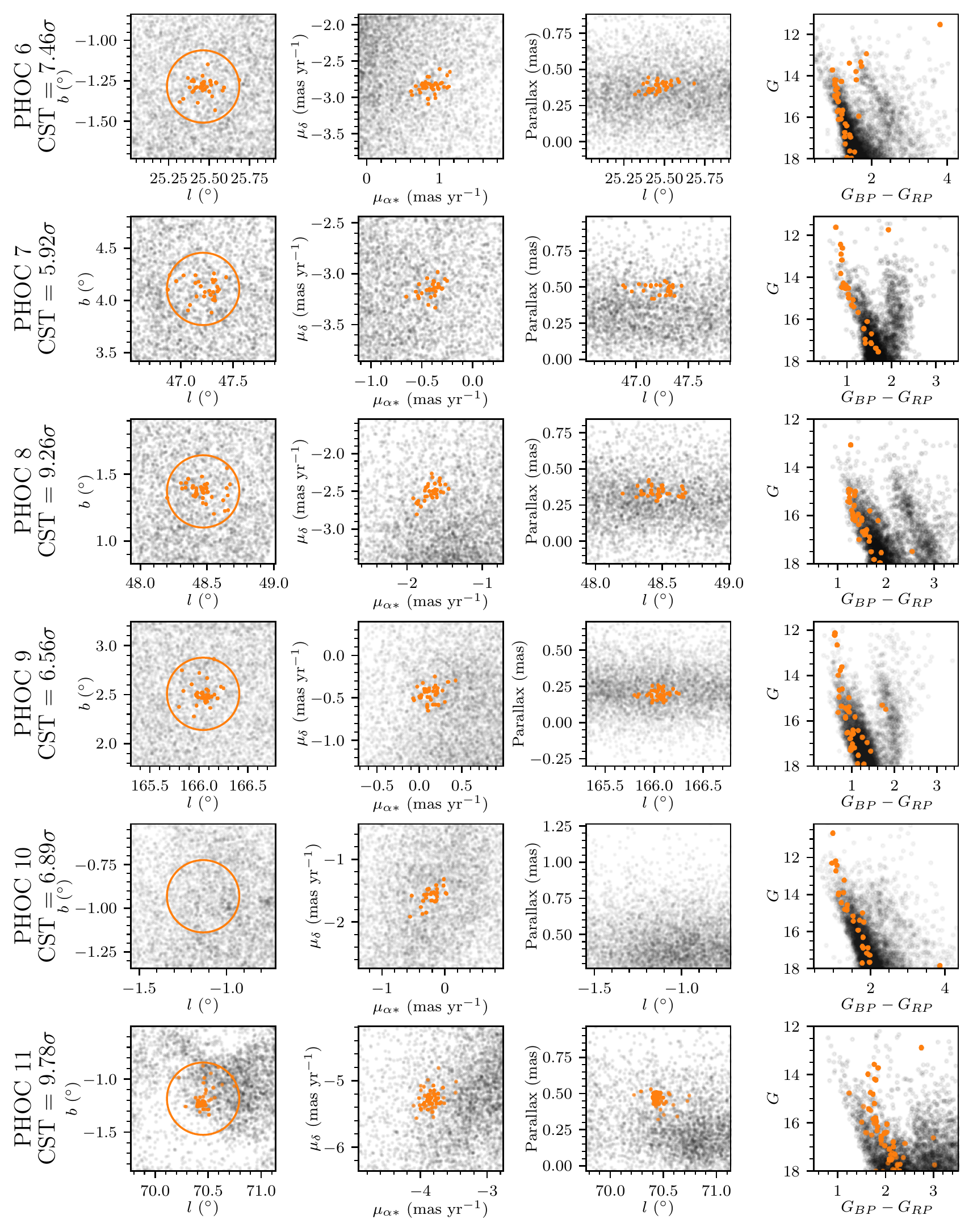}
   \caption{\reviewtwo{Plots of the new OCs PHOC 6 to 11, plotted in the same style as Fig.~\ref{fig:new_ocs_0}. }}%
\end{figure*}

\begin{figure*}[ht]
   \centering
   \includegraphics{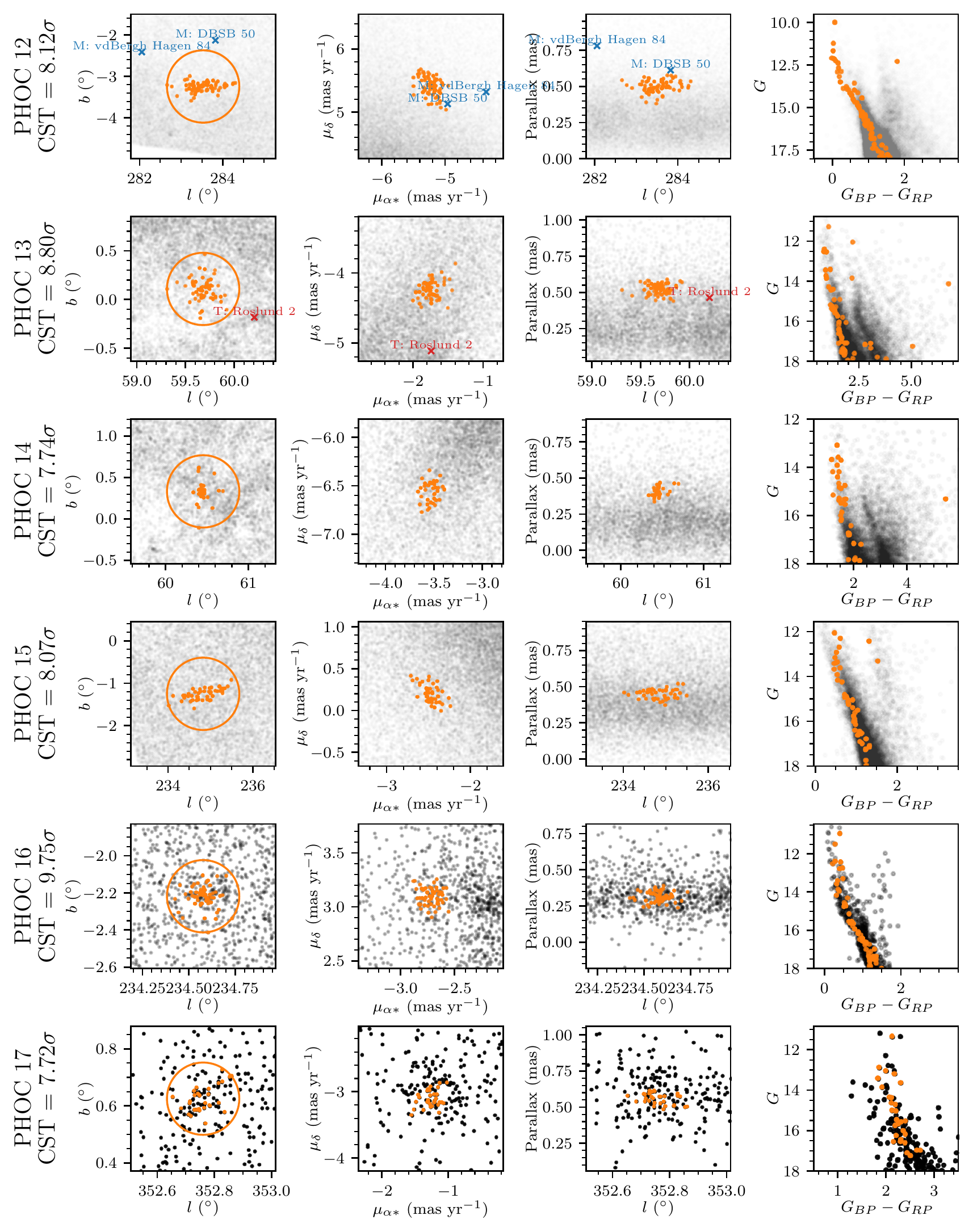}
   \caption{\reviewtwo{Plots of the new OCs PHOC 12 to 17, plotted in the same style as Fig.~\ref{fig:new_ocs_0}. }}%
\end{figure*}

\begin{figure*}[ht]
   \centering
   \includegraphics{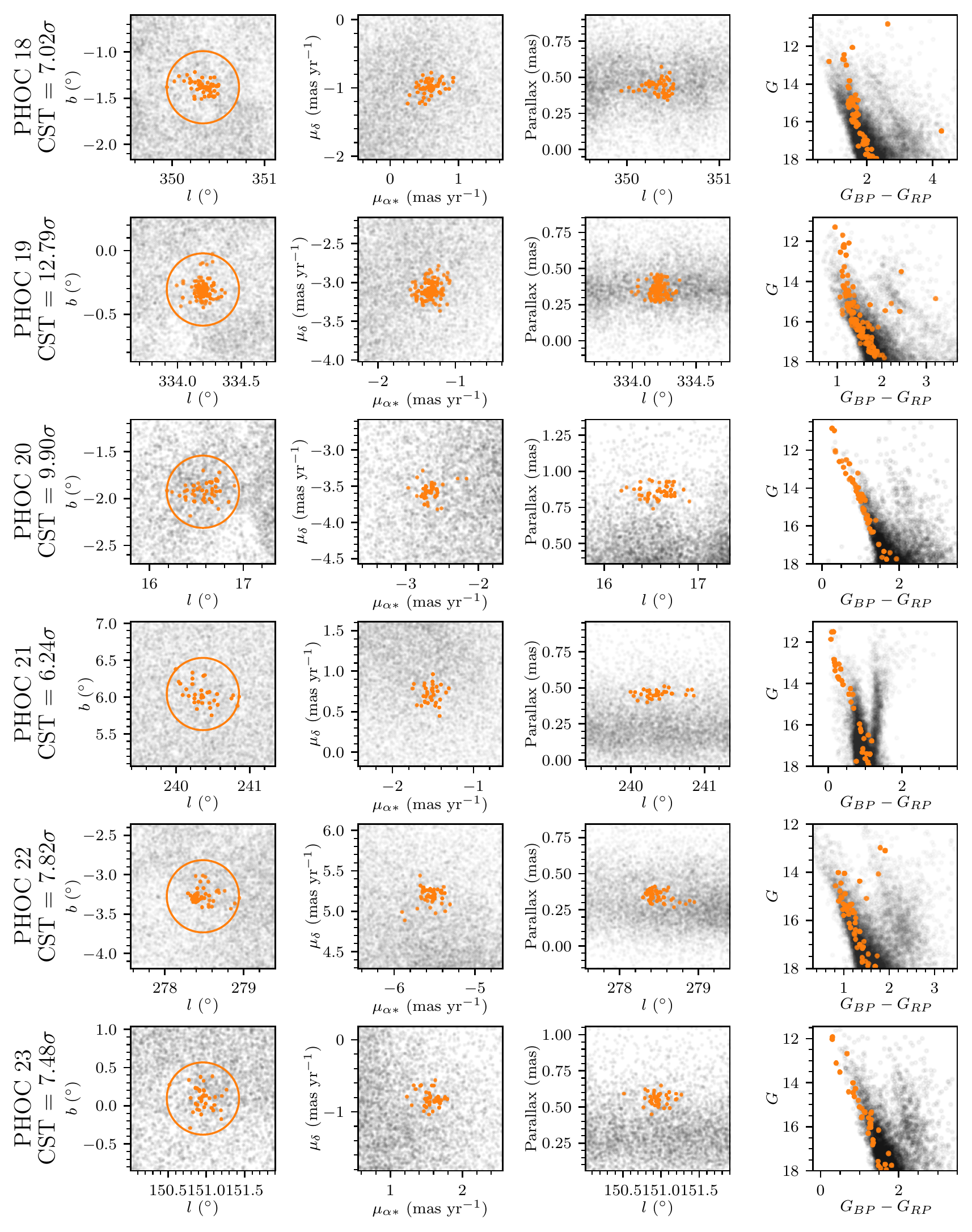}
   \caption{\reviewtwo{Plots of the new OCs PHOC 18 to 23, plotted in the same style as Fig.~\ref{fig:new_ocs_0}. }}%
\end{figure*}

\begin{figure*}[ht]
   \centering
   \includegraphics{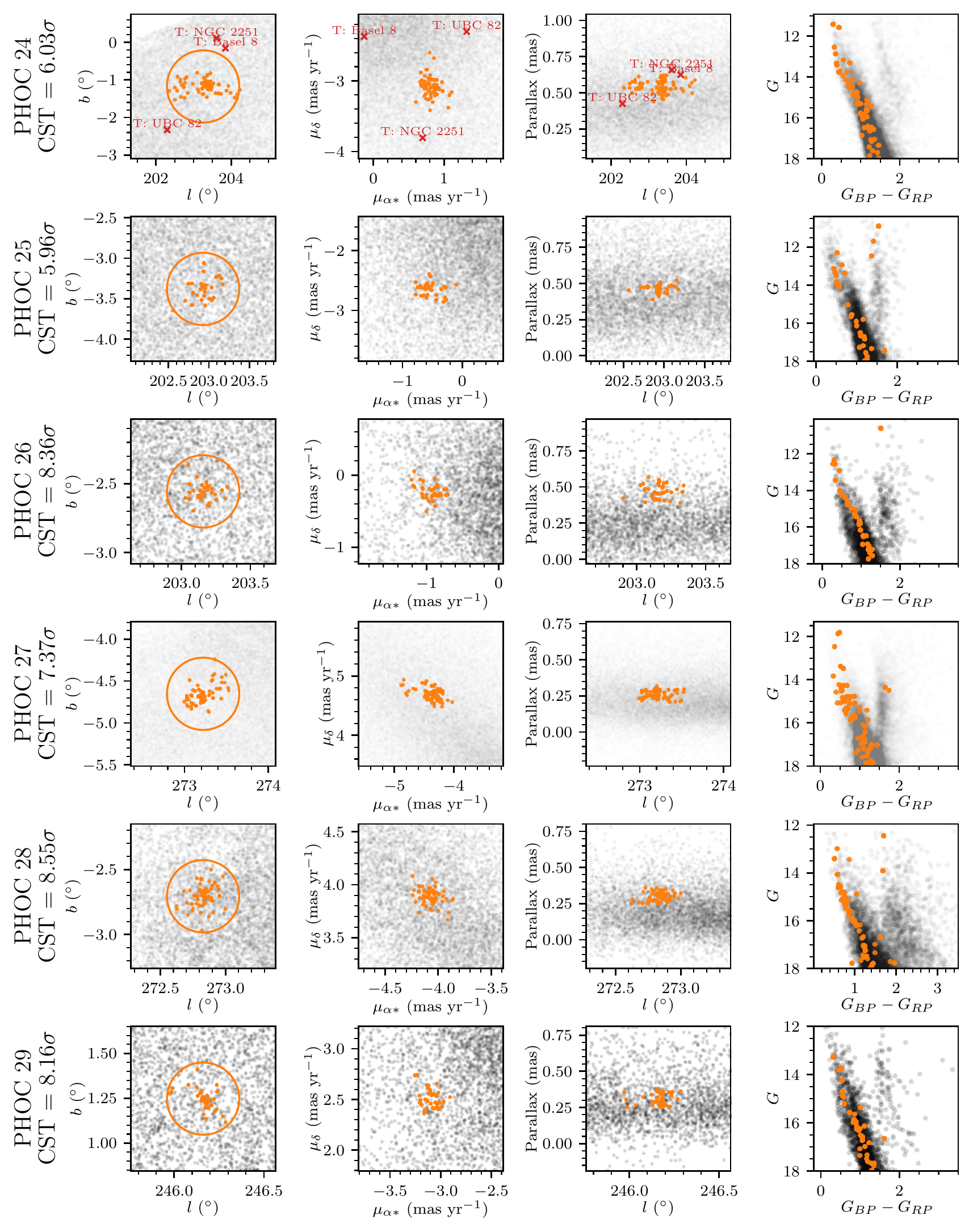}
   \caption{\reviewtwo{Plots of the new OCs PHOC 24 to 29, plotted in the same style as Fig.~\ref{fig:new_ocs_0}. }}%
\end{figure*}

\begin{figure*}[ht]
   \centering
   \includegraphics{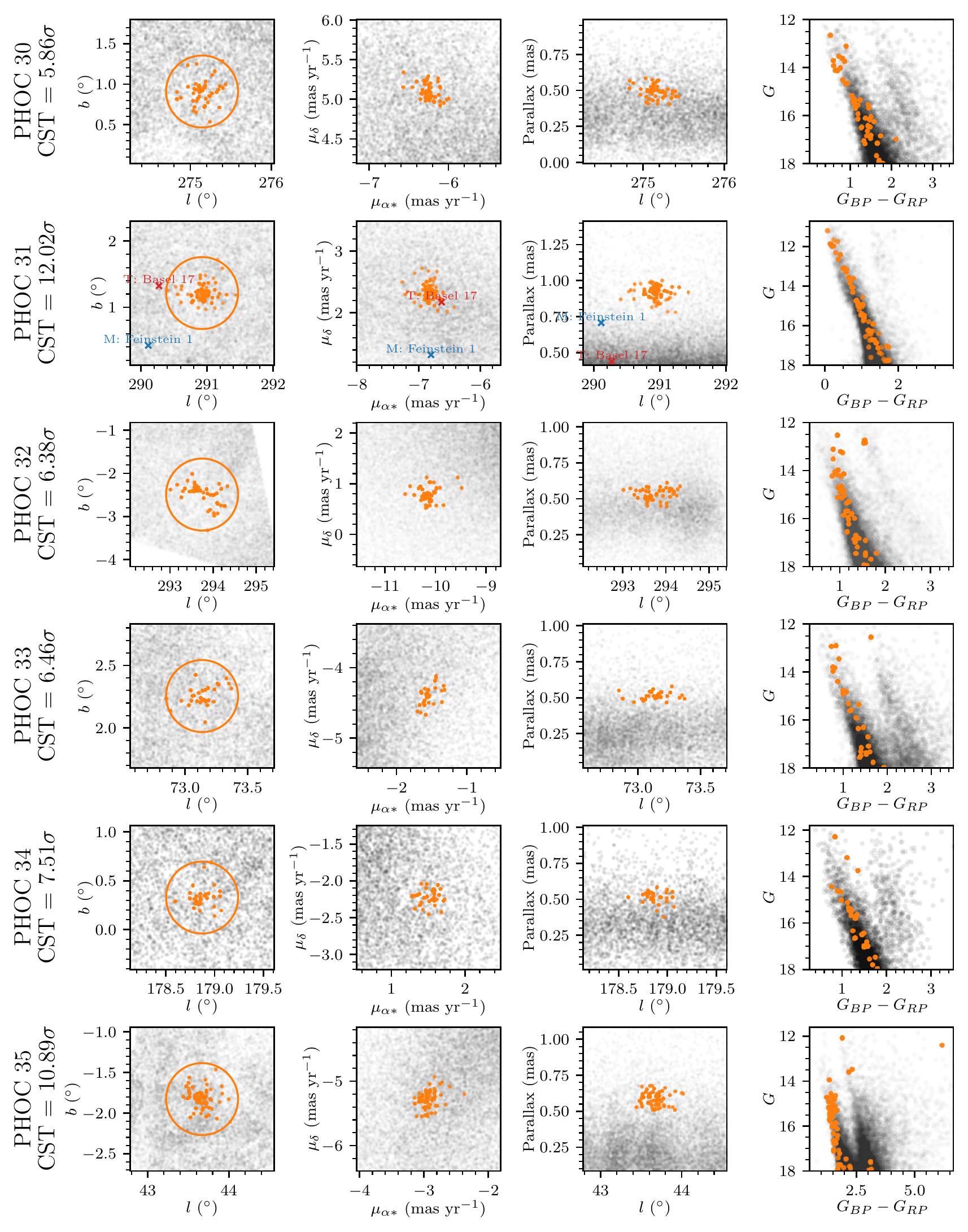}
   \caption{\reviewtwo{Plots of the new OCs PHOC 30 to 35, plotted in the same style as Fig.~\ref{fig:new_ocs_0}. }}%
\end{figure*}

\begin{figure*}[ht]
   \centering
   \includegraphics{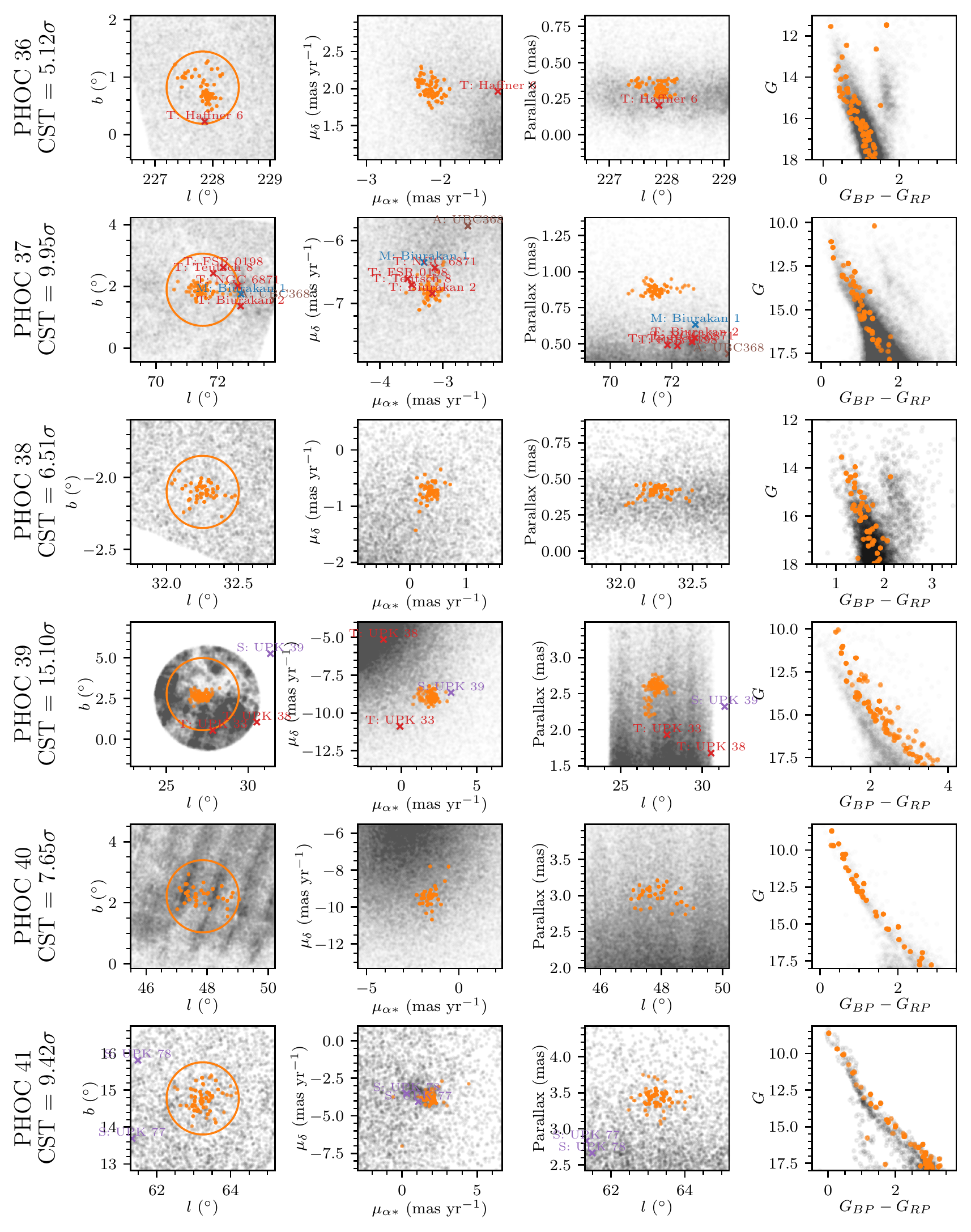}
   \caption{\reviewtwo{Plots of the new OCs PHOC 36 to 41, plotted in the same style as Fig.~\ref{fig:new_ocs_0}. }}%
\end{figure*}

%-------------------------------------------------------------------

\section{List of fields used in this study}\label{app:fields}
\begin{table*}

% Define first header
\caption{\label{table:sky_locations}Sky locations and HEALPix indices of the central pixels included in this study.}

\centering
\begin{tabular}{cccc|cccc}

\hline\hline
Number & $\alpha$ ($^\circ$) & $\delta$ ($^\circ$) & Level 5 HEALPix pixel\tablefootmark{a} & Number & $\alpha$ ($^\circ$) & $\delta$ ($^\circ$) & Level 5 HEALPix pixel\tablefootmark{a} \\
\hline

0  & 313.6 & -12.0 & 12238 & 50 & 318.1 & 46.6  & 3844  \\
1  & 104.1 & 18.2  & 5976  & 51 & 91.4  & 30.0  & 6106  \\
2  & 128.0 & -41.8 & 9817  & 52 & 136.9 & -54.3 & 9434  \\
3  & 343.9 & 69.4  & 3953  & 53 & 357.6 & 61.9  & 3575  \\
4  & 90.0  & 6.0   & 5900  & 54 & 48.1  & 46.6  & 772   \\
5  & 281.2 & -3.6  & 7564  & 55 & 76.5  & 45.0  & 365   \\
6  & 99.8  & 9.6   & 5909  & 56 & 120.9 & -31.4 & 9940  \\
7  & 92.8  & -6.0  & 5364  & 57 & 278.4 & -23.3 & 7243  \\
8  & 98.4  & 6.0   & 5563  & 58 & 293.9 & 22.0  & 3585  \\
9  & 281.2 & -25.9 & 7235  & 59 & 143.7 & -51.3 & 9437  \\
10 & 113.9 & -30.0 & 9946  & 60 & 34.4  & 64.9  & 915   \\
11 & 61.9  & 40.2  & 403   & 61 & 246.1 & -27.3 & 10738 \\
12 & 225.0 & -64.9 & 10432 & 62 & 274.2 & -17.0 & 7278  \\
13 & 106.9 & -8.4  & 5420  & 63 & 169.3 & -58.9 & 9484  \\
14 & 281.2 & -8.4  & 7552  & 64 & 84.4  & 31.4  & 6124  \\
15 & 73.1  & 31.4  & 284   & 65 & 168.2 & -61.9 & 9480  \\
16 & 286.9 & 13.2  & 7663  & 66 & 325.8 & 52.8  & 3860  \\
17 & 80.2  & 41.8  & 346   & 67 & 246.1 & -32.8 & 10702 \\
18 & 78.8  & 48.1  & 378   & 68 & 321.1 & 57.4  & 3870  \\
19 & 268.6 & -30.0 & 7205  & 69 & 302.3 & 35.7  & 3657  \\
20 & 303.8 & 34.2  & 3651  & 70 & 116.7 & -17.0 & 10158 \\
21 & 152.1 & -58.9 & 9340  & 71 & 88.6  & 27.3  & 6094  \\
22 & 120.9 & -10.8 & 5396  & 72 & 272.8 & -31.4 & 7192  \\
23 & 316.6 & 48.1  & 3846  & 73 & 85.8  & 30.0  & 6118  \\
24 & 295.3 & 23.3  & 3588  & 74 & 203.6 & -58.9 & 10427 \\
25 & 319.2 & 35.7  & 3317  & 75 & 48.6  & 52.8  & 792   \\
26 & 357.0 & 67.9  & 3927  & 76 & 315.0 & 46.6  & 3843  \\
27 & 112.5 & -20.7 & 9983  & 77 & 274.2 & 30.0  & 8153  \\
28 & 143.4 & -31.4 & 10004 & 78 & 112.5 & -18.2 & 5376  \\
29 & 272.8 & -18.2 & 7275  & 79 & 299.5 & 30.0  & 3606  \\
30 & 299.5 & 35.7  & 3658  & 80 & 289.7 & 10.8  & 7655  \\
31 & 136.6 & -48.1 & 9462  & 81 & 307.8 & 52.8  & 3875  \\
32 & 157.5 & -57.4 & 9506  & 82 & 98.4  & 23.3  & 6008  \\
33 & 257.3 & -38.7 & 10610 & 83 & 111.1 & -14.5 & 5385  \\
34 & 36.6  & 66.4  & 918   & 84 & 285.5 & 32.8  & 3630  \\
35 & 261.6 & -37.2 & 10612 & 85 & 255.9 & -37.2 & 10616 \\
36 & 204.8 & -60.4 & 10425 & 86 & 272.8 & -8.4  & 7387  \\
37 & 245.0 & -49.7 & 10543 & 87 & 300.9 & 34.2  & 3656  \\
38 & 258.8 & -37.2 & 10611 & 88 & 272.8 & -20.7 & 7272  \\
39 & 310.8 & 12.0  & 3117  & 89 & 23.8  & 64.9  & 911   \\
40 & 277.0 & -14.5 & 7290  & 90 & 102.7 & 0.0   & 5530  \\
41 & 122.3 & -22.0 & 10126 & 91 & 317.8 & 40.2  & 3496  \\
42 & 278.4 & 23.3  & 8056  & 92 & 109.7 & -31.4 & 9956  \\
43 & 105.5 & -19.5 & 5209  & 93 & 165.0 & -58.9 & 9483  \\
44 & 146.7 & -55.9 & 9428  & 94 & 95.6  & -10.8 & 5331  \\
45 & 91.4  & 19.5  & 5994  & 95 & 71.7  & 41.8  & 361   \\
46 & 61.7  & 49.7  & 439   & 96 & 253.1 & -34.2 & 10704 \\
47 & 319.2 & 38.7  & 3490  & 97 & 282.7 & 0.0   & 7578  \\
48 & 193.1 & -43.4 & 10904 & 98 & 95.6  & -23.3 & 5216  \\
49 & 97.0  & 7.2   & 5905  & 99 & 119.5 & -24.6 & 10122 \\

\hline

\end{tabular}

\tablefoot{
\tablefoottext{a}{To reproduce each full field, the eight nearest neighbour HEALPix level 5 pixels must also be selected.}
}

\end{table*}

\end{appendix}

\end{document}